\documentclass[12pt]{article}
\usepackage{amssymb,amsmath}

\usepackage{times,txfonts}

\usepackage{titling}
\usepackage{graphicx}
\usepackage[usenames]{color}
\usepackage[dvipsnames]{xcolor}
\usepackage{bm}
\usepackage{xspace}
\usepackage[paper=letterpaper,twoside,top=1.8cm,bottom=1.8cm,left=1.5cm,
  right=1.5cm,bindingoffset=0.0cm,voffset=0.0cm]{geometry}
\usepackage{hyperref}
\hypersetup{
  colorlinks   = true,
  citecolor    = blue,
  linkcolor    = blue,
  urlcolor     = blue
}
\usepackage{cite}

\DeclareFontFamily{OT1}{pzc}{}
\DeclareFontShape{OT1}{pzc}{m}{it}{<-> s * [1.2] pzcmi7t}{}
\DeclareMathAlphabet{\mathpzc}{OT1}{pzc}{m}{it}

\newcommand{\ham}{\mathpzc{H}}
\newcommand{\muB}{\mu_\textrm{B}}
\newcommand{\gmuH}{h_\textrm{red}}
\newcommand{\gmuHc}{h_{\textrm{red},\textrm{c}1}}
\newcommand{\Ncell}{N_\textrm{mag}}
\newcommand{\Hknot} {\mathpzc{H}^0_\textrm{\,LSW}}
\newcommand{\HknotE}{\mathpzc{H}^0_\textrm{\,LSW,even}}
\newcommand{\HknotO}{\mathpzc{H}^0_\textrm{\,LSW,odd}}
\newcommand{\HintSW}{\mathpzc{H}_{\,\textrm{NLSW}}}
\newcommand{\matHee}[1]{H_{\textrm{ee},#1}}
\newcommand{\matHoo}[1]{H_{\textrm{oo},#1}}
\newcommand{\matHeo}[1]{H_{\textrm{eo},#1}}
\newcommand{\MF}[1]{\langle{#1}\rangle^{}_{0}}
\renewcommand{\Re}[1]{{\mathrm{Re}\,{#1}}}

\newcommand{\ket}[1]{\bigl\lvert{#1}\bigr\rangle}
\newcommand{\bra}[1]{\bigl\langle{#1}\bigr\rvert}

\newcommand{\etal}{\textit{et al.}\xspace}
\newcommand{\ArticleType}{article\xspace}
\newcommand{\bsf}[1]{\textbf{#1}}
\newcommand{\estJ}{$J = 1.74$\,meV\xspace}
\newcommand{\estD}{$\Delta = 0.85$\xspace}
\newcommand{\estJc}{$J_c / J = 0.09$\xspace}
\newcommand{\estg}{$g_\perp = 3.95$\xspace}

\title{%
  {\bf
    The nature of spin excitations in the one-third magnetization plateau phase of Ba$_3$CoSb$_2$O$_9$    
  }
}%
\author{%
  Y. Kamiya,$^{1,\ast}$
  L. Ge,$^{2}$
  Tao Hong,$^{3}$
  Y. Qiu,$^{4}$
  D.~L. Quintero-Castro,$^{5}$
  Z. Lu,$^{5}$
  H.~B. Cao,$^{3}$
  \\
  M. Matsuda,$^{3}$
  E.~S.~Choi,$^{6}$
  C.~D. Batista,$^{7,8}$
  M. Mourigal,$^{2}$
  H.~D. Zhou,$^{6,7}$
  and
  J. Ma$^{9,10,7,\ast}$
  \vspace{6pt}
  \\
  \normalsize{$^1$\textit{Condensed Matter Theory Laboratory, RIKEN, Wako, Saitama 351-0198, Japan}}
  \\
  \normalsize{$^2$\textit{School of Physics, Georgia Institute of Technology, Atlanta, GA 30332, USA}}
  \\
  \normalsize{$^3$\textit{Neutron Scattering Division, Oak Ridge National Laboratory, Oak Ridge, Tennessee 37831, USA}}
  \\
  \normalsize{$^4$\textit{NIST Centre for Neutron Research, National Institute of Standards and Technology,}} \\
  \normalsize{\textit{Gaithersburg, MD 20899, USA}}
  \\ 
  \normalsize{$^5$\textit{Helmholtz-Zentrum Berlin f\"ur Materialien und Energie, D-14109 Berlin, Germany}}
  \\
  \normalsize{$^6$\textit{National High Magnetic Field Laboratory, Florida State University, Tallahassee, FL 32310, USA}}
  \\
  \normalsize{$^7$\textit{Department of Physics and Astronomy, University of Tennessee, Knoxville, Tennessee 37996, USA}}
  \\
  \normalsize{$^8$\textit{Neutron Scattering Division and Shull-Wollan Center, Oak Ridge National Laboratory,}} \\
  \normalsize{\textit{Oak Ridge, Tennessee 37831, USA}}
  \\ 
  \normalsize{$^9$\textit{Key Laboratory of Artificial Structures and Quantum Control, Department of Physics and Astronomy,}} \\
  \normalsize{\textit{Shanghai Jiao Tong University, Shanghai 200240, China}}
  \\
  \normalsize{$^{10}$\textit{Collaborative Innovation Center of Advanced Microstructures, Nanjing, Jiangsu 210093, China}}
}%

\begin{document}
\let\emph\textit
\maketitle
\baselineskip23pt

\textbf{%
  Magnetization plateaus in quantum magnets---where bosonic quasiparticles crystallize into emergent spin superlattices---are spectacular yet simple examples of collective quantum phenomena escaping classical description. While magnetization plateaus have been observed in a number of spin-1/2 antiferromagnets, the description of their magnetic excitations remains an open theoretical and experimental challenge. Here, we investigate the dynamical properties of the triangular-lattice spin-1/2 antiferromagnet Ba$_3$CoSb$_2$O$_9$ in its one-third magnetization plateau phase using a combination of nonlinear spin-wave theory and neutron scattering measurements. The agreement between our theoretical treatment and the experimental data demonstrates that magnons behave semiclassically in the plateau in spite of the purely quantum origin of the underlying magnetic structure. This allows for a quantitative determination of Ba$_3$CoSb$_2$O$_9$ exchange parameters. We discuss the implication of our results to the deviations from semiclassical behavior observed in zero-field spin dynamics of the same material and conclude they must have an intrinsic origin.
}%

\clearpage
\baselineskip18pt

Quantum fluctuations favor collinear spin order in frustrated magnets~\cite{Villain1980,Shender1982,Henley1989}, which can be qualitatively different from the classical limit ($S \to \infty$)~\cite{Lacroix2011}. In particular, quantum effects can produce \emph{magnetization plateaus}~\cite{Chubukov1991,Zhitomirsky2000,Hida2005,Lacroix2011C10,Seabra2016,Ye2017}, where the magnetization is pinned at a fraction of its saturation value. Magnetization plateaus can be interpreted as crystalline states of bosonic particles, and are naturally stabilized by easy-axis exchange anisotropy, which acts as strong off-site repulsion~\cite{Rice2002,Giamarchi2008,Zapf2014,Haravifard2016}. However, the situation is less evident and more intriguing for \emph{isotropic} Heisenberg magnets, which typically have no plateaus in the classical limit. In a seminal work, Chubukov and Golosov predicted the 1/3 magnetization plateau in the quantum triangular lattice Heisenberg antiferromagnet (TLHAFM), corresponding to an up-up-down (UUD) state~\cite{Chubukov1991}. Their predictions were confirmed by numerical studies~\cite{Zhitomirsky2000,Honecker2004,Farnell2009,Sakai2011,Chen2013,Yamamoto2014,Yamamoto2015,Nakano2017} and extended to plateaus in other models~\cite{Zhitomirsky2000}. Experimentally, the 1/3 plateau has been observed in the spin-1/2 isosceles triangular lattice material Cs$_2$CuBr$_4$,~\cite{Ono2003,Ono2004,Ono2005,Tsujii2007,Fortune2009} as well as in the equilateral triangular lattice materials RbFe(MoO$_4$)$_2$ ($S=5/2$)~\cite{Inami1996,Svistov2003,White2013} and Ba$_3$CoSb$_2$O$_9$ (effective $S= 1/2$)~\cite{Doi2004,Shirata2012,Zhou2012,Susuki2013,Naruse2014,Koutroulakis2015,Quirion2015,Ma2016,Sera2016,Ito2017}.

Notwithstanding the progress in the search of quantum plateaus, much less is known about their \emph{excitation spectra}. Given that they are stabilized by quantum fluctuations, it is natural to ask if these fluctuations strongly affect the excitation spectrum. The qualitative difference between the plateau and the classical orderings may appear to invalidate spin-wave theory. For instance, the UUD state in the equilateral TLHAFM is not a classical ground state unless the magnetic field H is fine-tuned~\cite{Alicea2009}. Consequently, a naive spin wave treatment is doomed to instability. On the other hand, spin wave theory builds on the assumption of an ordered moment $\lvert\langle{\mathbf{S}_\mathbf{r}}\rangle\rvert$ close to the full moment. Given that a sizable reduction of $\lvert\langle{\mathbf{S}_\mathbf{r}}\rangle\rvert$ is unlikely within the plateau because of the gapped nature of the spectrum, a spin wave description could be adequate. Although this may seem in conflict with the order-by-disorder mechanism~\cite{Villain1980,Shender1982,Henley1989} stabilizing the plateau~\cite{Alicea2009,Coletta2013,Coletta2016}, this phenomenon is produced by the zero-point energy correction $E_{\rm zp} = (1/2)\sum_{\mathbf{q}}\omega_\mathbf{q} + O(S^0)$ ($\omega_\mathbf{q}$ is the spin wave dispersion), which does not necessarily produce a large moment size reduction.

One of our goals is to resolve this seemingly contradictory situation. Recently, Alicea \etal developed a method to fix the unphysical spin-wave instability~\cite{Alicea2009}. This proposal awaits experimental verification because the excitation spectrum has not been measured over the entire Brillouin zone for any fluctuation-induced plateau. We will demonstrate that the modified nonlinear spin wave (NLSW) approach indeed reproduces the magnetic excitation spectrum of Ba$_3$CoSb$_2$O$_9$ within the 1/3 plateau~\cite{Doi2004,Shirata2012,Zhou2012,Susuki2013,Naruse2014,Koutroulakis2015,Quirion2015,Ma2016,Sera2016,Ito2017}. The resulting model parameters confirm that the anomalous \emph{zero-field} dynamics reported in two independent experiments~\cite{Ma2016,Ito2017} must be intrinsic and non-classical.

\begin{figure}[t]
  \centering
  \includegraphics[width=0.85\hsize, bb=0 0 928 317]{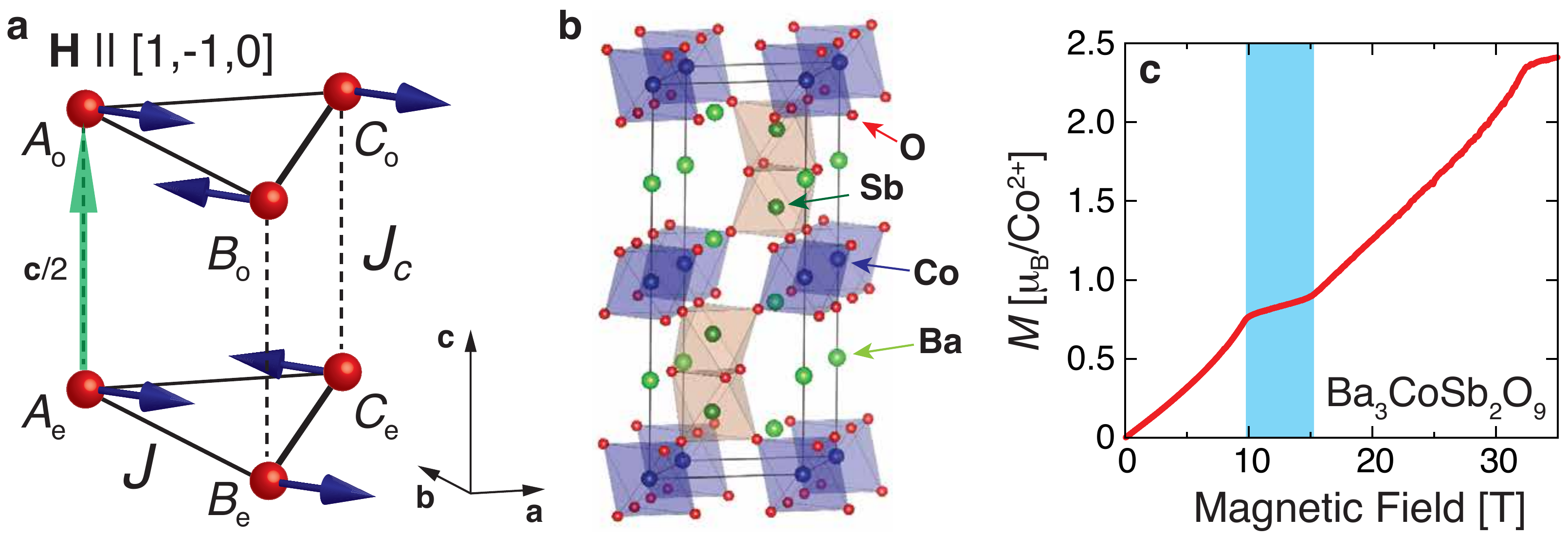}
  \vspace{10pt}
  \caption{%
    \label{fig:UUD}
    \bsf{Stacked triangular lattice and the UUD state.}
    (\bsf{a}) Spin structure in the quasi-2D lattice.
    (\bsf{b}) Crystal structure of Ba$_3$CoSb$_2$O$_9$.
    (\bsf{c}) Magnetization curve for $\mathbf{H} \parallel ab$ at $T = 0.6$\,K highlighting the 1/3 plateau (the finite slope is due to Van Vleck paramagnetism~\cite{Susuki2013}).
  }%
\end{figure}

In this \ArticleType, we present a comprehensive study of magnon excitations in the 1/3 magnetization plateau phase of a quasi-two-dimensional (quasi-2D) TLHAFM with easy-plane exchange anisotropy. Our study combines NLSW theory with in-field inelastic neutron scattering (INS) measurements of Ba$_3$CoSb$_2$O$_9$. The Hamiltonian is
\begin{align}
  \ham &= 
  J \sum_{\langle{\mathbf{r}\mathbf{r}'}\rangle} \left( S^x_{\mathbf{r}} S^x_{\mathbf{r}'} + S^y_{\mathbf{r}} S^y_{\mathbf{r}'} + \Delta S^z_{\mathbf{r}} S^z_{\mathbf{r}'} \right)
  \notag\\
  &
  + J_c \sum_{\mathbf{r}} \left( S^x_{\mathbf{r}} S^x_{\mathbf{r}+\frac{\mathbf{c}}{2}} + S^y_{\mathbf{r}} S^y_{\mathbf{r}+\frac{\mathbf{c}}{2}} + \Delta S^z_{\mathbf{r}} S^z_{\mathbf{r}+\frac{\mathbf{c}}{2}} \right)
  - \gmuH \sum_{\mathbf{r}} S_{\mathbf{r}}^x,
  \label{eq:model}
\end{align}
where $\langle{\mathbf{r}\mathbf{r}'}\rangle$ runs over in-plane nearest-neighbor (NN) sites of the stacked triangular lattice and $\frac{\mathbf{c}}{2}$ corresponds to the interlayer spacing (Fig.~\ref{fig:UUD}a). $J$ ($J_c$) is the antiferromagnetic intralayer (interlayer) NN exchange and $0 \le \Delta < 1$. The magnetic field is in the in-plane ($x$) direction (we use a spin-space coordinate frame where $x$ and $y$ are in the $ab$ plane and $z$ is parallel to $c$). $\gmuH = g_\perp\muB H$ is the reduced field and $g_\perp$ is the in-plane $g$-tensor component.

This model describes Ba$_3$CoSb$_2$O$_9$ (Fig.~\ref{fig:UUD}b), which comprises triangular layers of effective spin 1/2 moments arising from the $\mathcal{J} = 1/2$ Kramers doublet of Co$^{2+}$ in a perfect octahedral ligand field. Excited multiplets are separated by a gap of 200--300\,K due to spin-orbit coupling, which is much larger than the N\'eel temperature $T_\text{N} = 3.8$\,K. Below $T = T_\text{N}$, the material develops conventional 120$^\circ$ ordering with wavevector $\mathbf{Q} = (1/3,1/3,1)$~\cite{Doi2004}. Experiments confirmed a 1/3 magnetization plateau for $\mathbf{H} \!\parallel\! ab$ (Fig.~\ref{fig:UUD}c)~\cite{Shirata2012,Susuki2013,Koutroulakis2015,Quirion2015,Sera2016}, which is robust down to the lowest temperatures. We compute the dynamical spin structure factor using NLSW theory in the 1/3 plateau phase. We also provide neutron diffraction evidence of the UUD state within the 1/3 plateau of Ba$_3$CoSb$_2$O$_9$, along with maps of the excitation spectrum obtained from INS. The excellent agreement between theory and experiment demonstrates the semiclassical nature of magnons within the 1/3 plateau phase, despite the quantum fluctuation-induced nature of the ground state ordering.

\section*{Results}
\subsection*{Quantum-mechanical stabilization of the plateau in quasi-2D TLHAFMs.}
While experimental observations show that deviations from the ideal 2D TLHAFM are small in Ba$_3$CoSb$_2$O$_9$ \cite{Susuki2013,Ma2016,Ito2017}, a simple variational analysis shows that any $J_c > 0$ is enough to destabilize the UUD state \emph{classically}. Thus, a naive spin wave treatment leads to instability for $J_c > 0$. However, the gapped nature of the spectrum~\cite{Chubukov1991} implies that this phase must have a finite range of stability in quasi-2D materials. This situation must be quite generic among fluctuation-induced plateaus, as they normally require special conditions to be a classical ground state~\cite{Kawamura1985,Seabra2016,Ye2017}.

\begin{figure}[t]
  \centering
  \includegraphics[width=\hsize, bb=0 0 1321 556]{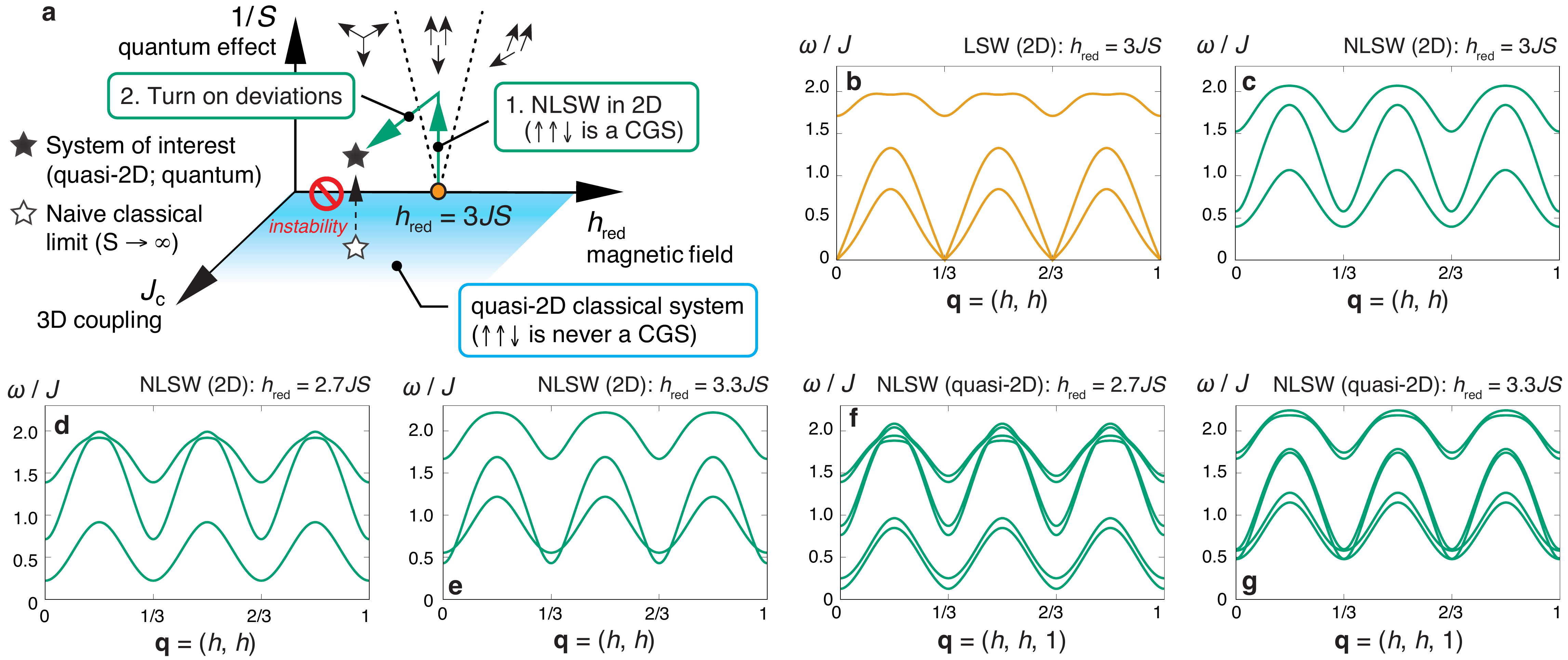}
  \caption{%
    \label{fig:NLSW}
    \bsf{Scheme of NLSW theory for the 1/3 magnetization plateau in the quasi-2D TLHAFM.}
    (\bsf{a}) Illustration of the procedure used  to compute the spectrum.
    The filled and open star symbols represent the target quasi-2D quantum system and its naive classical limit, respectively. 
    The spin structures favored by quantum fluctuation
    are shown on the $J_c = 0$ plane.
    (\bsf{b}) LSW spectrum along the high-symmetry direction of the Brillouin zone evaluated for $S = 1/2$, $J_c = 0$, $\Delta = 0.85$ and $\gmuH = 3 J S$, where the UUD state is a classical ground state (CGS).
    (\bsf{c}--\bsf{e}) NLSW spectra for $J_c = 0$ and $\Delta = 0.85$ with (\bsf{c}) $\gmuH = 3 J S$, (\bsf{d}) $\gmuH = 2.7 J S$, and (\bsf{e}) $\gmuH = 3.3 J S$.
    (\bsf{f}, \bsf{g}) NLSW spectra for $J_c / J = 0.09$ and $\Delta = 0.85$ with (\bsf{f}) $\gmuH = 2.7 J S$ and (\bsf{g}) $\gmuH = 3.3 J S$.
  }%
\end{figure}

To put this into a proper semiclassical framework, we apply Alicea \etal's trick originally applied to a distorted triangular lattice~\cite{Alicea2009}. Basically, we make a ``detour'' in the parameter space with the additional $1/S$-axis quantifying the quantum effect (Fig.~\ref{fig:NLSW}). Instead of expanding the Hamiltonian around $S \to \infty$ for the actual model parameters, we start from the special point, $J_c = 0$, $\gmuH = 3 J S$, and a given value of $0 \le \Delta \le 1$, that includes the UUD state in its classical ground state manifold. Assuming the spin structure in Fig.~\ref{fig:UUD}a, we define
\begin{align}
  S^x_\mathbf{r} = \tilde{S}^z_\mathbf{r},~~
  S^y_\mathbf{r} = \tilde{S}^y_\mathbf{r},~~
  S^z_\mathbf{r} = -\tilde{S}^x_\mathbf{r},
  \label{eq:rotation:1}
\end{align}
for $\mathbf{r} \in$ $A_\mathrm{e}$, $B_\mathrm{e}$, $A_\mathrm{o}$, and $C_\mathrm{o}$ and
\begin{align}
  S^x_\mathbf{r} = -\tilde{S}^z_\mathbf{r},~~
  S^y_\mathbf{r} = \tilde{S}^y_\mathbf{r},~~
  S^z_\mathbf{r} = \tilde{S}^x_\mathbf{r},
  \label{eq:rotation:2}
\end{align}
for $\mathbf{r} \in C_\mathrm{e}, B_\mathrm{o}$. Introducing the Holstein-Primakoff bosons, $a^{(\dag)}_{\mu,\mathbf{r}}$, with $1 \le \mu \le 6$ being the sublattice index for $A_\mathrm{e}$, $B_\mathrm{e}$, $C_\mathrm{e}$, $A_\mathrm{o}$, $B_\mathrm{o}$, and $C_\mathrm{o}$ in this order, we have
\begin{align}
  \tilde{S}^z_\mathbf{r} &= S - a^{\dag}_{\mu,\mathbf{r}} a^{\;}_{\mu,\mathbf{r}},
  \notag\\
  \tilde{S}^+_\mathbf{r}
  &= \tilde{S}^x_\mathbf{r} + i \tilde{S}^y_\mathbf{r}
  \approx \sqrt{2S} \left( 1 - \frac{a^{\dag}_{\mu,\mathbf{r}} a^{\;}_{\mu,\mathbf{r}}}{4S} \right) a^{\;}_{\mu,\mathbf{r}},
  \label{eq:HP}
\end{align}
and $\tilde{S}^-_\mathbf{r}$ $=$ $\bigl( \tilde{S}^+_\mathbf{r} \bigr)^\dag$ for $\mathbf{r}\in\mu$, truncating higher order terms irrelevant for the quartic interaction. We evaluate magnon self-energies arising from decoupling of the quartic term.

\begin{figure}[t]
  \centering
  \includegraphics[width=0.95\hsize, bb=0 0 898 652]{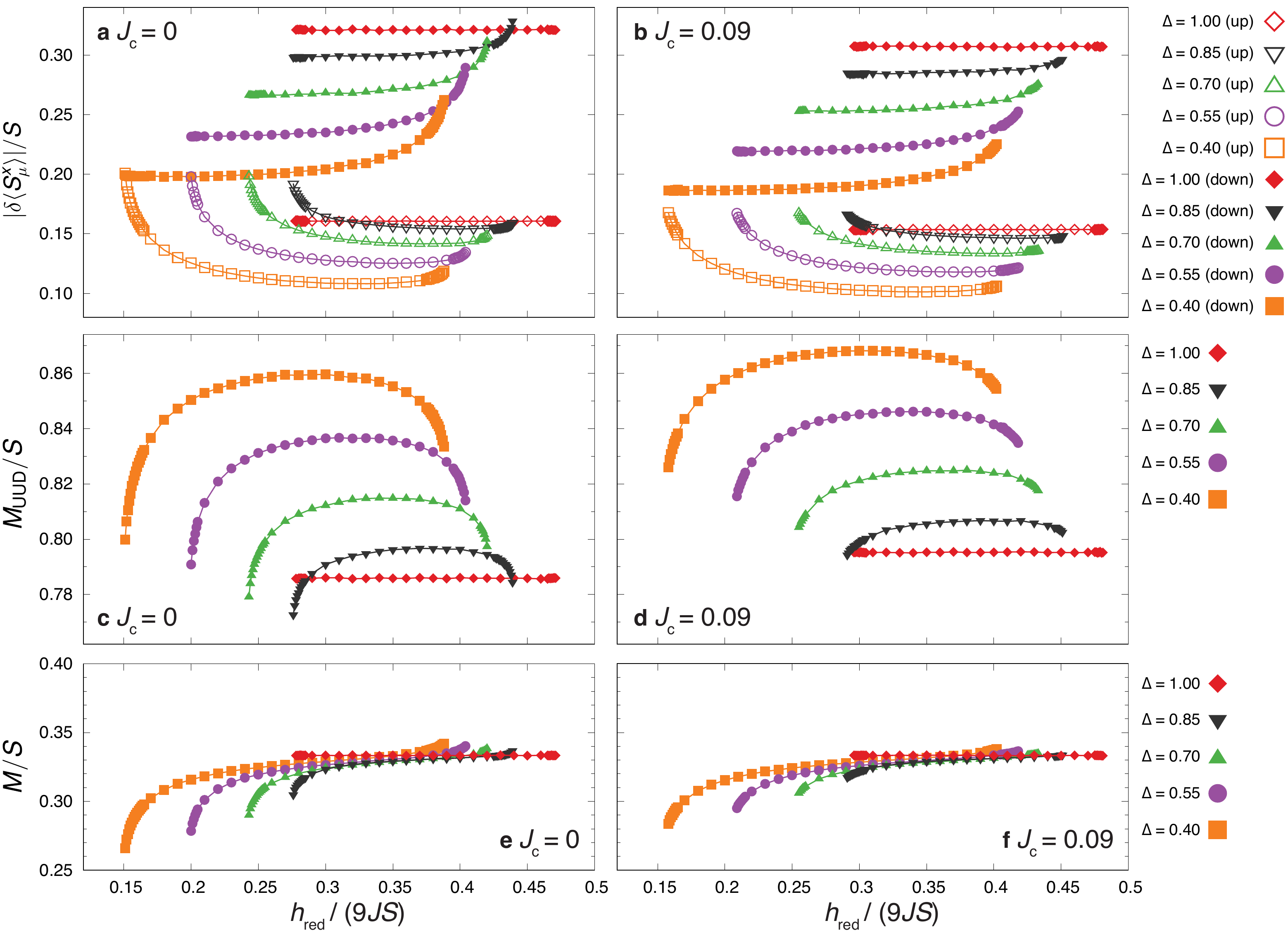}
  \caption{%
    \label{fig:Muud}
    \bsf{Calculated ordered moment within the 1/3 plateau for spin 1/2.}
    Top row: magnitude of the reduction, $\lvert{\delta\langle{S^x_{\mu}}\rangle}\rvert$, of the sublattice ordered moments (normalized by $S$) with the designated moment directions (up or down) for (\bsf{a}) $J_c = 0$ and (\bsf{b}) $J_c = 0.09J$; for sublattices with up spins, we average $\delta\langle{S^x_{\mu}}\rangle$ over the corresponding two sublattices, discarding the small variance that appears for $J_c > 0$.
    Middle row: normalized staggered magnetization $M_\mathrm{UUD}/S$ [Eq.~\eqref{eq:Muud}] for (\bsf{c}) $J_c = 0$ and (\bsf{d}) $J_c = 0.09J$.
    Bottom row: normalized (uniform) magnetization $M/S$ for (\bsf{e}) $J_c = 0$ and (\bsf{f}) $J_c = 0.09J$. The results correspond to the local stability range of the plateau (with the precision $0.001 \times 9JS$ for $\gmuH$).
  }%
\end{figure}

As shown in Fig.~\ref{fig:NLSW}b, the linear spin wave (LSW) spectrum for $J_c = 0$ and $\gmuH = 3 J S$ features two $q$-linear gapless branches at $\mathbf{q} = 0$, both of which are gapped out by the magnon-magnon interaction (Fig.~\ref{fig:NLSW}c). Small deviations from $J_c = 0$ and $\gmuH = 3 J S$ do not affect the local stability of the UUD state because the gap must close continuously. Thus, we can investigate the excitation spectrum of quasi-2D systems for fields near $\gmuH = 3 J S$. Figures~\ref{fig:NLSW}d and \ref{fig:NLSW}e show the spectra for $\gmuH$ shifted by $\pm 10\%$ from $\gmuH = 3 J S$, where we still keep $J_c = 0$. For $\gmuH < 3 J S$, a band-touching and subsequent hybridization appear between the middle and the top bands around $\mathbf{q}=(1/6,1/6)$ (Fig.~\ref{fig:NLSW}d). For $\gmuH > 3 J S$, a level-crossing between the middle and bottom bands appears at around $\mathbf{q}=0$ (Fig.~\ref{fig:NLSW}e). A small $J_c > 0$ splits the three branches into six (Figs.~\ref{fig:NLSW}f and \ref{fig:NLSW}g). Figures~\ref{fig:Muud}a and \ref{fig:Muud}b  show the reduction of the sublattice ordered moments for $S = 1/2$, $J_c = 0$, $0.09J$, and selected values of $\Delta$. We find $\lvert{\delta\langle{S^x_{\mu}}\rangle}\rvert/S \lesssim$~30\% throughout the local stability range of the plateau. Thus, our semiclassical approach is fully justified within the plateau phase. Figures~\ref{fig:Muud}c and \ref{fig:Muud}d show the field dependence of the staggered magnetization,
\begin{align}
  M_\mathrm{UUD} = \frac{1}{6}\left(\langle{S^x_{A_\mathrm{e}}}\rangle + \langle{S^x_{B_\mathrm{e}}}\rangle - \langle{S^x_{C_\mathrm{e}}}\rangle + \langle{S^x_{A_\mathrm{o}}}\rangle - \langle{S^x_{B_\mathrm{o}}}\rangle + \langle{S^x_{C_\mathrm{o}}}\rangle\right),
  \label{eq:Muud}
\end{align}
which is almost field-independent; a slightly enhanced field-independence appears for small $\Delta$. Similarly, while the magnetization is not conserved for $\Delta \ne 1$, it is nearly pinned at 1/3 for the most part of the plateau (Figs.~\ref{fig:Muud}e and \ref{fig:Muud}f).

\begin{figure}[t]
  \centering
  \includegraphics[width=0.9\hsize, bb=0 0 871 252]{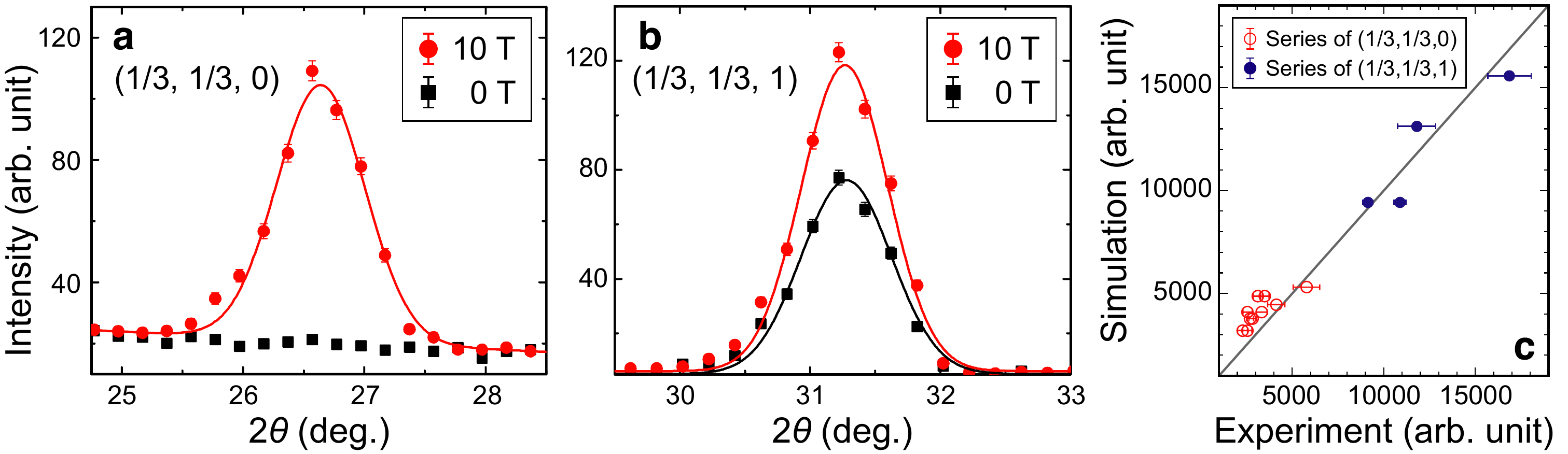}
  \caption{%
    \label{fig:ND}
    \bsf{Neutron diffraction data for Ba$_3$CoSb$_2$O$_9$ at 0\,T and 10\,T.}
    (\bsf{a}) $\mathbf{q} = (1/3, 1/3, 0)$ and
    (\bsf{b}) $\mathbf{q} = (1/3, 1/3, 1)$ measured at $T = 1.5$\,K.
    (\bsf{c}) Comparison of the diffraction intensities between the experiment and the simulation at $T = 1.5$\,K and $\mu_0 H = 10$\,T (the solid line is a guide to the eye).
    Error bars represent one standard deviation.
  }%
\end{figure}

\subsection*{UUD state in Ba$_3$CoSb$_2$O$_9$.}
Next we show experimental evidence for the UUD state in Ba$_3$CoSb$_2$O$_9$ by neutron diffraction measurements within the plateau phase for field applied along the [1,-1,0] direction. We used the same single crystals reported in Refs.~\cite{Ma2016,Zhou2012}, grown by the traveling-solvent floating-zone technique and characterized by neutron diffraction, magnetic susceptibility, and heat capacity measurements. The space group is $P$6$_3$/$mmc$, with the lattice constants $a= b =$ 5.8562~\AA\xspace and $c =$ 14.4561~\AA. The site-disorder between Co$^{2+}$ and Sb$^{5+}$ is negligible with the standard deviation of 1\%, as reported elsewhere~\cite{Ma2016}. The magnetic and structural properties are consistent with previous reports and confirm high quality of the crystals~\cite{Doi2004,Shirata2012,Zhou2012,Susuki2013,Naruse2014,Koutroulakis2015,Quirion2015,Ma2016,Sera2016,Ito2017}. These crystals were oriented in the $(h,h,l)$ scattering plane. The magnetic Bragg peaks at $(1/3, 1/3, 0)$ and $(1/3, 1/3, 1)$ were measured at $T = 1.5$\,K (Figs.~\ref{fig:ND}a and \ref{fig:ND}b). The large intensity at both $(1/3, 1/3, 0)$ and $(1/3, 1/3, 1)$ confirms the UUD state at $\mu_0 H \ge 9.8$\,T~\cite{Zhou2012} (Fig.~\ref{fig:ND}c). The estimated ordered moment is 1.65(3)\,$\muB$ at 10\,T and 1.80(9)\,$\muB$ at 10.9\,T. They correspond to 85(2)\% and 93(5)\% of the full moment~\cite{Susuki2013}, roughly coinciding with the predicted range (Fig.~\ref{fig:Muud}d). This diffraction pattern can be contrasted with that of the 120$^{\circ}$ state, characterized by a combination of the large intenisty at $(1/3, 1/3, 1)$ and lack of one at $(1/3, 1/3, 0)$. Our diffraction result is fully consistent with previous nuclear magnetic resonance (NMR)~\cite{Koutroulakis2015} and magnetization measurements~\cite{Shirata2012,Susuki2013}.

\begin{figure*}[t]
  \centering
  \includegraphics[width=\hsize,bb=0 0 755 376,clip]{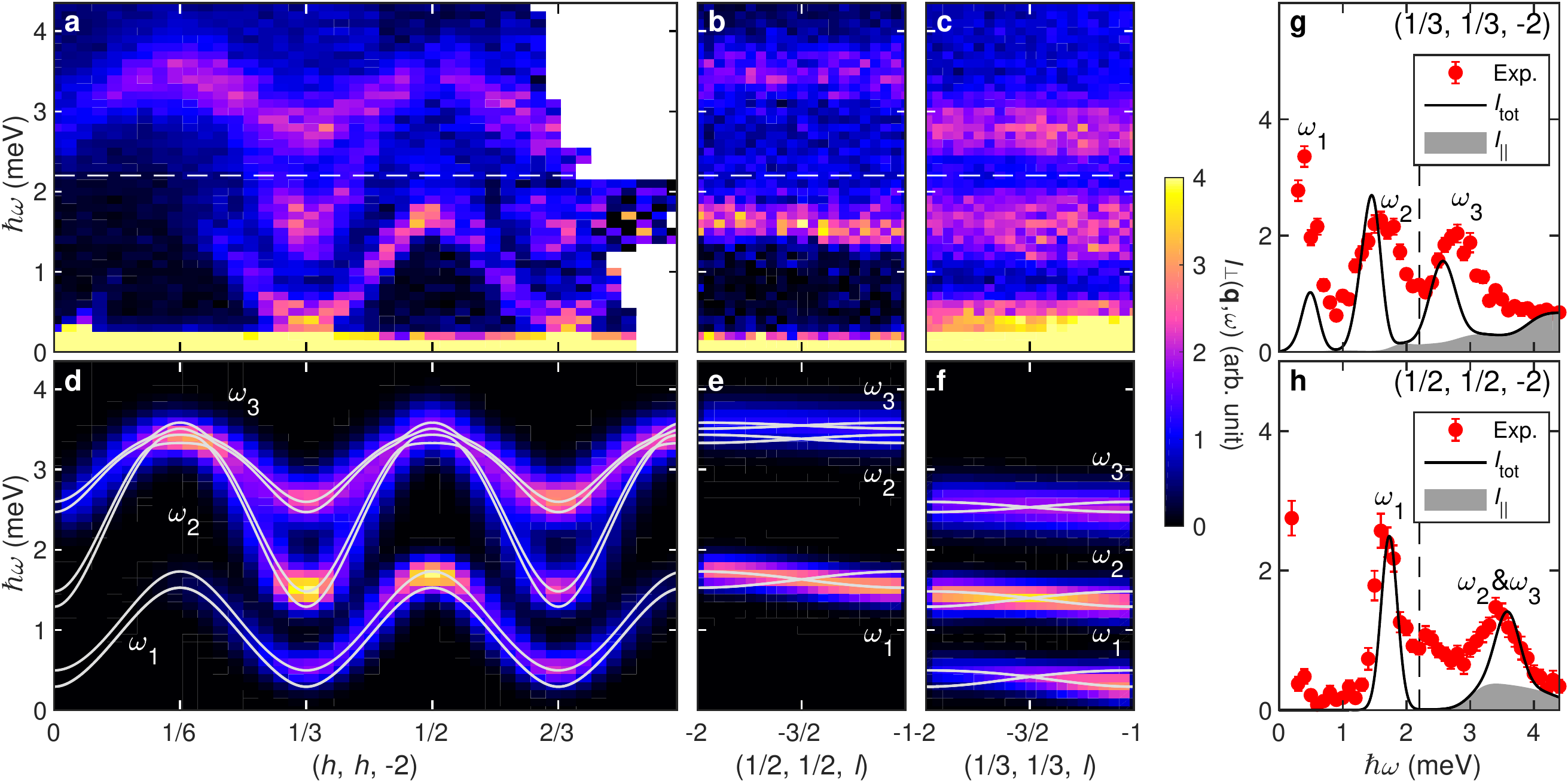}
  \caption{%
    \label{fig:Sqw}
    \bsf{Excitation spectrum in the UUD phase of Ba$_3$CoSb$_2$O$_9$.}
    (\bsf{a})--(\bsf{c}) Experimental scattering intensity at $\mu_0 H = 10.5$\,T and $T = 0.5$\,K with momentum transfer (\bsf{a}) $\mathbf{q} = (h, h, -2)$, (\bsf{b}) $\mathbf{q} = (1/2, 1/2, l)$, and (\bsf{c}) $\mathbf{q} = (1/3, 1/3, l)$. 
    (\bsf{d})--(\bsf{f}) Calculated transverse part of the scattering intensity, $I_{\perp}(\mathbf{q},\omega)$, obtained by NLSW theory along the same momentum cuts as in (\bsf{a})--(\bsf{c}) for \estJ, \estD, \estJc, and \estg. The solid lines show the magnon poles. 
    (\bsf{g}) and (\bsf{h}) Energy dependence of the calculated scattering intensity, $I_\mathrm{tot}(\mathbf{q},\omega)$ (solid line), compared with the experiment for (\bsf{g}) $\mathbf{q} = (1/3,1/3,-2)$ and (\bsf{h}) $\mathbf{q}=(1/2,1/2,-2)$ (error bars represent one standard deviation). The longitudinal contribution to the scattering intensity, $I_{\parallel}(\mathbf{q},\omega)$, is plotted separately as a shaded area. The energy of the outgoing neutrons is $E_{\rm f}=5$\,meV (3\,meV) above (below) the dashed line in (\bsf{a})--(\bsf{c}), (\bsf{g}), and (\bsf{h}).
  }%
\end{figure*}

\begin{figure}[t]
  \centering
  \includegraphics[width=0.6\hsize,bb=0 0 514 561]{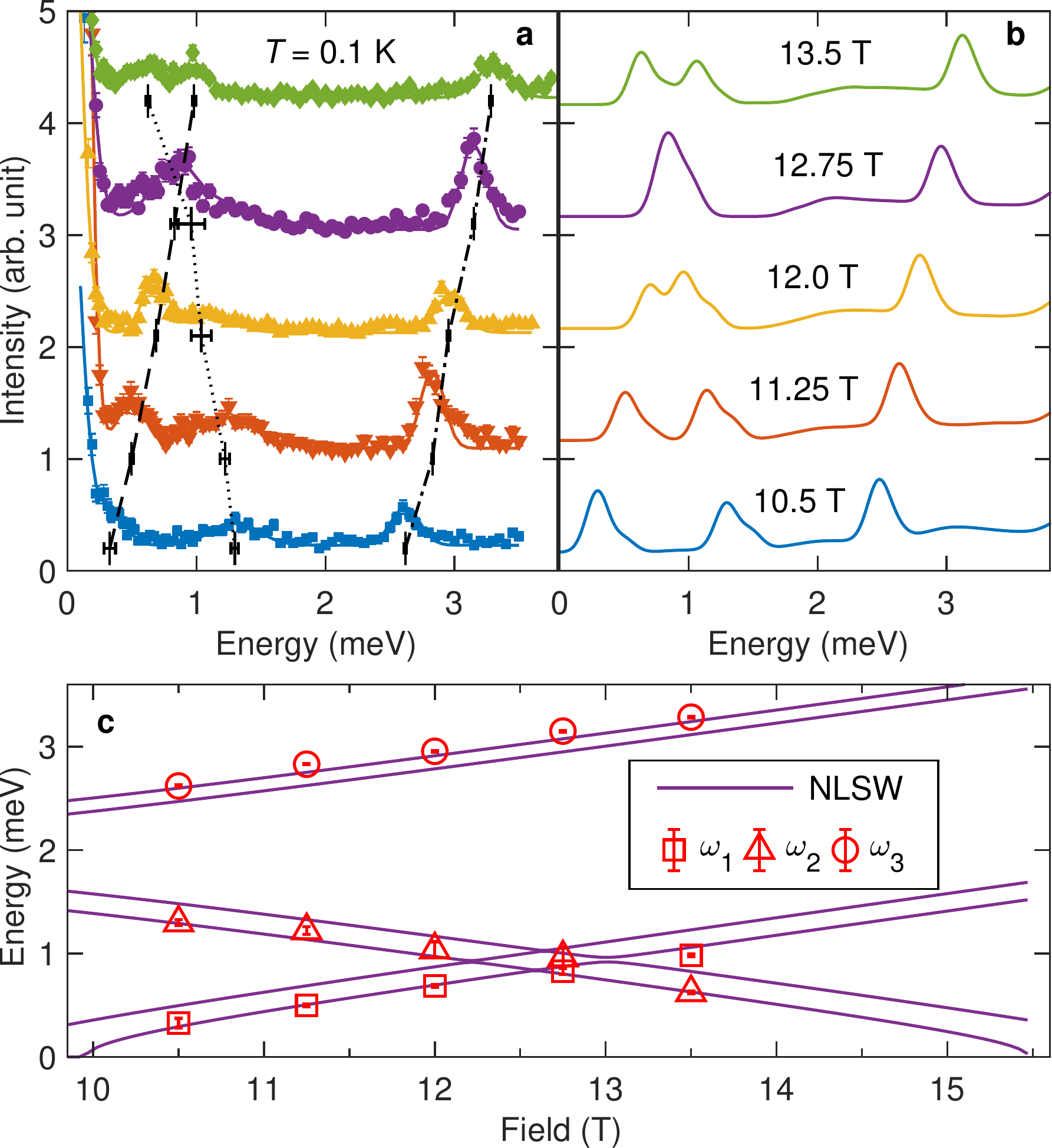}
  \caption{%
    \label{fig:Hdep}  
    \bsf{Field-dependence of $\omega_1$--$\omega_3$ at $\mathbf{q = (1/3, 1/3, 1)}$ in the UUD phase of Ba$_3$CoSb$_2$O$_9$.}
    (\bsf{a}) Constant-$q$ scans for the selected values of the magnetic field at $T = 0.1$\,K. The solid-lines show the fitting to Gaussian functions. The locations of the magnon peaks are indicated, where the solid ($\omega_1$), dashed ($\omega_2$), and dotted ($\omega_3$) lines are guides to the eye.
    (\bsf{b}) Simulation of the line-shape by NLSW theory for \estJ, \estD, \estJc, and \estg convoluted with the assumed resolution 0.2\,meV.
    (\bsf{c}) Comparison of the fitted magnon frequencies $\omega_1$--$\omega_3$ against the NLSW poles.
    Error bars represent one standard deviation.
  }%
\end{figure}

\subsection*{Excitation spectrum.}
We now turn to the dynamical properties in the UUD phase. Figures~\ref{fig:Sqw}a--\ref{fig:Sqw}c show the INS intensity $I(\mathbf{q},\omega) \equiv k_{\rm i}/k_{\rm f} (\mathrm{d}^2\sigma/\mathrm{d}\Omega\mathrm{d}E_{\rm f})$ along high-symmetry directions. The applied magnetic field $\mu_0 H = 10.5$\,T is relatively close to the transition field $\mu_0 H_{\mathrm{c}1} = 9.8$\,T~\cite{Zhou2012} bordering on the low-field coplanar ordered phase~\cite{Koutroulakis2015}, while the temperature $T = 0.5$\,K is low enough compared to $T_{\textrm{N}} \approx 5$\,K~\cite{Quirion2015} for the UUD phase at this magnetic field. The in-plane dispersion shown in Fig.~\ref{fig:Sqw}a comprises a seemingly gapless branch at $\mathbf{q}\!=\!(1/3, 1/3, -1)$ (Fig.~\ref{fig:Sqw}c), and two gapped modes centered around 1.6\,meV and 2.7\,meV. Due to the interlayer coupling, each mode corresponds to two non-degenerate branches. As their splitting is below the instrumental resolution, we simply refer to them as $\omega_1$, $\omega_2$ and $\omega_3$, unless otherwise mentioned (Fig.~\ref{fig:Sqw}). The dispersions along the $c$-direction are nearly flat, as shown in Figs.~\ref{fig:Sqw}b and \ref{fig:Sqw}c for ${\bf q}\!=\!(1/2, 1/2, l)$ and ${\bf q}\!=\!(1/3, 1/3, l)$, respectively, reflecting the quasi-2D lattice~\cite{Susuki2013,Ma2016}. Among the spin wave modes along $\mathbf{q}\!=\!(1/2, 1/2, l)$ and $\mathbf{q}\!=\!(1/3, 1/3, l)$ in Figs.~\ref{fig:Sqw}b and \ref{fig:Sqw}c, $\omega_1$
for $\mathbf{q}\!=\!(1/2, 1/2, l)$
displays a relatively sharp spectral line. As discussed below, most of the broadening stems from the different intensities of the split modes due to finite $J_c$.

Comparing the experiment against the NLSW calculation, we find that the features of the in-plane spectrum in Fig.~\ref{fig:Sqw}a are roughly captured by the theoretical calculation near the low-field onset of the plateau in Fig.~\ref{fig:NLSW}f ($\gmuH = 2.7 J S \approx 1.03\, \gmuHc$). This observation is in accord with the fact that the applied field ($\mu_0 H = 10.5$\,T) is close to $\mu_0 H_{\mathrm{c}1} = 9.8$\,T~\cite{Zhou2012}. To refine the quantitative comparison, we calculate the scattering intensity $I_{\mathrm{tot}}(\mathbf{q},\omega) \equiv (\gamma r_0/2)^2 \left| F(\mathbf{q}) \right|^2 \sum_\alpha (1 - \hat{q}^\alpha \hat{q}^\alpha) g^2_{\alpha} \mathcal{S}^{\alpha\alpha}(\mathbf{q},\omega)$ where $F(\mathbf{q})$ denotes the magnetic form factor of Co$^{2+}$ corrected with the orbital contribution, $(\gamma r_0/2)^2$ is a constant, $\hat{q}^\alpha = q^\alpha/\left|\mathbf{q}\right|$, and $\mathcal{S}^{\alpha\alpha}(\mathbf{q},\omega)$ are the diagonal components of the dynamical structure factor evaluated at 10.5\,T; off-diagonal components are zero due to symmetry. Defining the UUD order as shown in Fig.~\ref{fig:UUD}a, transverse spin fluctuations related to single-magnon excitations appear in $\mathcal{S}^{yy}$ and $\mathcal{S}^{zz}$, while longitudinal spin fluctuations corresponding to the two-magnon continuum appear in the inelastic part of $\mathcal{S}^{xx}$, denoted as $\mathcal{S}_{\parallel}$. Accordingly, $I_{\mathrm{tot}}(\mathbf{q},\omega)$ can be separated into transverse $I_{\perp}$ and longitudinal $I_{\parallel}$ contributions. To compare with our experiments, the theoretical intensity is convoluted with momentum binning effects (only for $I_{\perp}$) and empirical instrumental energy resolution. Figures~\ref{fig:Sqw}d--\ref{fig:Sqw}f show the calculated $I_{\perp}(\mathbf{q},\omega)$, along the same high-symmetry paths as the experimental results in Figs.~\ref{fig:Sqw}a--\ref{fig:Sqw}c, for \estJ, \estD, \estJc, and \estg. The agreement between theory and experiment is excellent. When deriving these estimates, $J$ is controlled by the saturation field $\mu_0 H_\mathrm{sat} = 32.8$\,T for $\mathbf{H} \parallel \hat{c}$~\cite{Susuki2013}. To obtain the best fit, we also analyzed the field dependence of $\omega_1$, $\omega_2$ and $\omega_3$ (Fig.~\ref{fig:Hdep}).

Remarkably, the calculation in Figs.~\ref{fig:Sqw}d--\ref{fig:Sqw}f reproduces the dispersions almost quantitatively. It predicts a gapped $\omega_1$ mode, although the gap is below experimental resolution. The smallness of the gap is simply due to proximity to $H_{\mathrm{c}1}$. For each $\omega_i$, the band splitting due to $J_c$ yields pairs of poles $\omega^\pm_i$ dispersing with a phase difference of $\pi$ in the out-of-triangular-plane direction (Figs.~\ref{fig:Sqw}e and \ref{fig:Sqw}f). For each pair, however, one pole has a vanishing intensity for ${\bf q}\!=\!(1/2, 1/2, l)$. Consequently, $\omega_1$ along this direction is free from any extrinsic broadening caused by overlapping branches (Fig.~\ref{fig:Sqw}e), yielding a relatively sharp spectral line (Fig.~\ref{fig:Sqw}h). The corresponding bandwidth $\approx 0.2$\,meV (Fig.~\ref{fig:Sqw}b) provides a correct estimate for $J_c$. By contrast, for ${\bf q}\!=\!(1/3, 1/3, l)$, all six $\omega^\pm_i$ branches have non-zero intensity, which leads to broadened spectra and less obvious dispersion along $l$ (Figs.~\ref{fig:Sqw}c and \ref{fig:Sqw}f).

The field-dependence of $\omega_{1}$--$\omega_{3}$ at $\mathbf{q}\!=\! (1/3, 1/3, 1)$ is extracted from constant-$q$ scans at $T = 0.1$\,K for selected fields $10.5$\,T--$13.5$\,T within the plateau (Fig.~\ref{fig:Hdep}a). By fitting the field-dependence of the low-energy branches of $\omega_{1,2}$, which become gapless at a plateau edge, we obtain the quoted model parameters. The field dependence is reproduced fairly well (Figs.~\ref{fig:Hdep}b and \ref{fig:Hdep}c), although the calculation slightly underestimates $\omega_3$. We find $\omega_1$ and $\omega_3$ ($\omega_2$) increase (decreases) almost linearly in $H$, while the $\omega_1$ and $\omega_2$ branches cross around 12.6 T. The softening of $\omega_1$ ($\omega_2$) at the lower (higher) transition field induces the Y-like (V-like) state, respectively~\cite{Koutroulakis2015,Yamamoto2015}. The nonlinearity of the first excitation gap near these transitions (visible only in the calculation) is due to the anisotropy; there is no U(1) symmetry along the field direction for $\Delta \ne 1$.

\section*{Discussion}
Our work has mapped out the excitation spectrum in the 1/3 plateau---a manifestation of quantum order-by-disorder effect---in Ba$_3$CoSb$_2$O$_9$. Despite the quantum-mechanical origin of the ground state ordering, we have unambiguously demonstrated the \emph{semiclassical nature} of magnons in this phase. In fact, the calculated reduction of the sublattice magnetization, $\delta S_\mu = S - \lvert\langle{\mathbf{S}_{\mathbf{r}}}\rangle\rvert$ with $\mathbf{r}\in\mu$, is relatively small (Fig.~\ref{fig:Muud}b): $\delta S_{A_\mathrm{e}} = \delta S_{A_\mathrm{o}} = 0.083$, $\delta S_{B_\mathrm{e}} = \delta S_{C_\mathrm{o}} = 0.073$, and $\delta S_{C_\mathrm{e}} = \delta S_{B_\mathrm{o}} = 0.14$ at 10.5\,T for the quoted model parameters. This is consistent with the very weak intensity of the two-magnon continuum (Figs.~\ref{fig:Sqw}g and \ref{fig:Sqw}h). This semiclassical behavior is protected by the excitation gap induced by anharmonicity of the spin waves (magnon-magnon interaction). We note that a perfect collinear magnetic order does not break any continuous symmetry even for $\Delta = 1$, i.e., there is no gapless Nambu-Goldstone mode. The collinearity also means that three-magnon processes are not allowed~\cite{Zhitomirsky2013}. The gap is robust against perturbations, such as anisotropies, lattice deformations~\cite{Alicea2009}, or biquadratic couplings for $S > 1/2$ (a ferroquadrupolar coupling can stabilize the plateau even classically~\cite{Chubukov1991}). Thus, we expect the semiclassical nature of the excitation spectrum to be common to other 2D and quasi-2D realizations of fluctuation-induced plateaus, such as the 1/3 plateau in the spin-5/2 material RbFe(MoO$_4$)$_2$~\cite{White2013}. Meanwhile, it will be interesting to examine the validity of the semiclassical approach in quasi-1D TLHAFMs, such as Cs$_2$CuBr$_4$~\cite{Ono2004}, where quantum fluctuations are expected to be stronger.

Finally, we discuss the implications of our results for the \emph{zero-field} dynamical properties of the same material, where recent experiments revealed unexpected phenomena, such as broadening of the magnon peaks \emph{indescribable} by conventional spin-wave theory, large intensity of the high-energy continuum~\cite{Ma2016}, and the extension thereof to anomalously high frequencies~\cite{Ito2017}. Specifically, Ref.~\cite{Ma2016} reported magnon spectral-line broadened throughout the entire Brillouin zone, significantly beyond instrumental resolution, and a high frequency ($\gtrsim 2\,\textrm{meV}$) excitation continuum with an almost comparable spectral weight as single-magnon modes. All of these experimental observations indicate strong quantum effects. Given that the spin Hamiltonian has been reliably determined from our study of the plateau phase, it is interesting to reexamine if a semiclassical treatment of this Hamiltonian can account for the zero-field anomalies.

\begin{figure}[t]
  \centering
  \includegraphics[width=\hsize, bb=0 0 826 435]{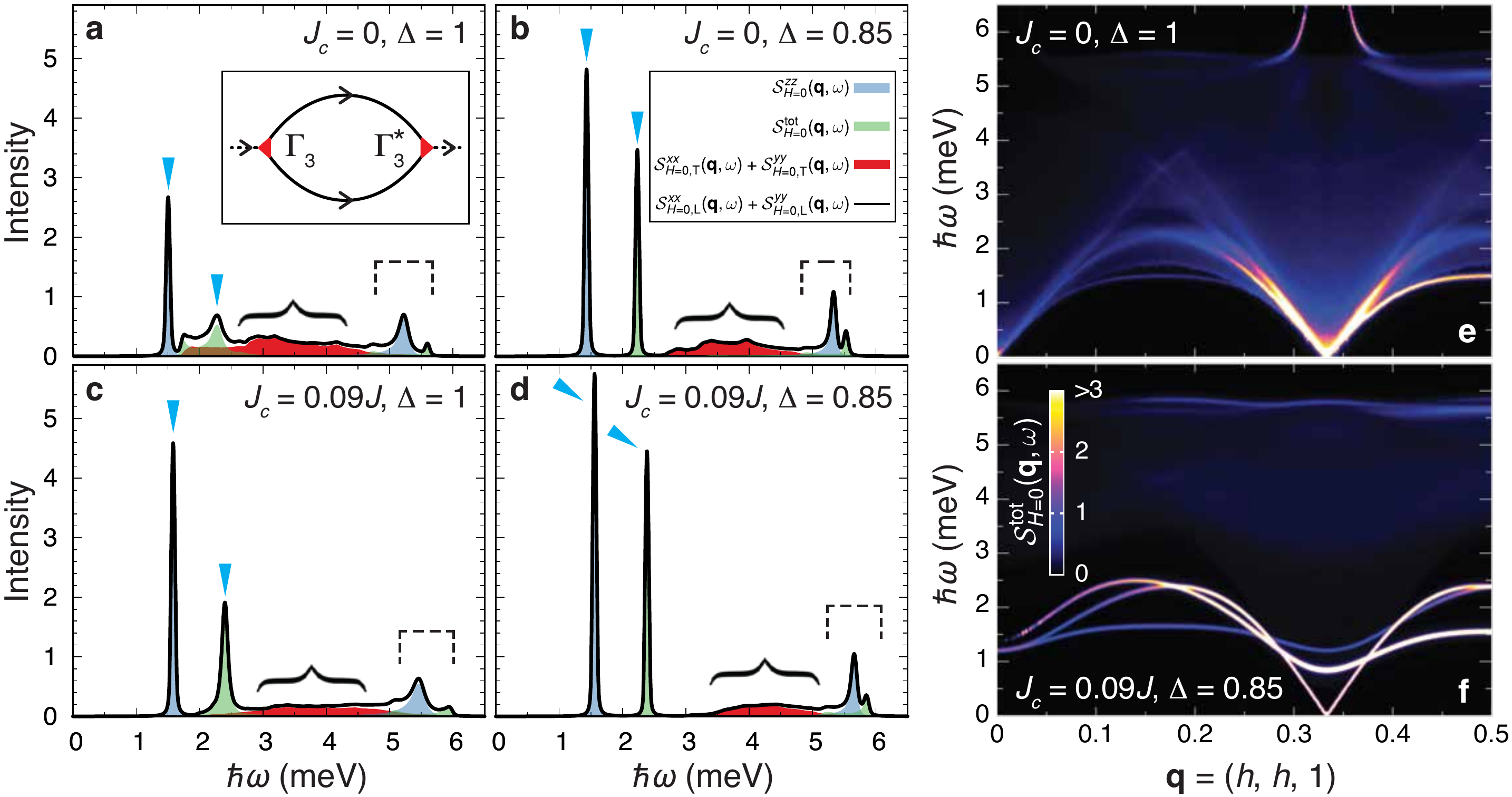}
  \caption{%
    \label{fig:120deg}
    \bsf{Calculated dynamical spin structure factor for the in-plane 120$^{\circ}$ state at $H = 0$ evaluated with NLSW theory for spin 1/2.}
    (\bsf{a})--(\bsf{d}) 
    The results of the frequency dependence at $\mathbf{q} = (1/2,1/2,1)$ (M point) for 
    (\bsf{a}) $J_c = 0$ and $\Delta = 1$ (the ideal TLHAFM),
    (\bsf{b}) $J_c = 0$ and $\Delta = 0.85$,
    (\bsf{c}) $J_c = 0.09J$ and $\Delta = 1$, and
    (\bsf{d}) $J_c = 0.09J$ and $\Delta = 0.85$.
    The results are convoluted with the energy resolution $0.015J$. The total spin structure factor $\mathcal{S}_{H=0}^{\mathrm{tot}}(\mathbf{q},\omega)$ (solid line) is divided into different components; $\mathcal{S}_{H=0}^{zz}(\mathbf{q},\omega)$ and $\mathcal{S}_{H=0,\, \mathrm{T}}^{xx}(\mathbf{q},\omega) + \mathcal{S}_{H=0,\,\mathrm{T}}^{yy}(\mathbf{q},\omega)$ are single-magnon contributions (``T'' denotes the transverse part), while the longitudinal (L) part $\mathcal{S}_{H=0,\,\mathrm{L}}^{xx}(\mathbf{q},\omega) + \mathcal{S}_{H=0,\,\mathrm{L}}^{yy}(\mathbf{q},\omega)$ corresponds to the two-magnon continuum; single magnon peaks (two-magnon continua) are indicated by arrows (curly brackets), whereas the dashed square brackets indicate anti-bonding single-magnon contributions, which are expected to be broadened by higher-order effect in $1/S$~\cite{Mourigal2013a}. The inset shows the lowest-order, $O(S^0)$, magnon self-energy incorporating the decay process of a single magnon into two magnons.
    (\bsf{e}) and (\bsf{f}) Intensity plots of $\mathcal{S}_{H=0}^{\mathrm{tot}}(\mathbf{q},\omega)$ along the high-symmetry direction in the Brillouin zone for (\bsf{e}) $J_c = 0$ and $\Delta = 1$ and (\bsf{f}) $J_c = 0.09J$ and $\Delta = 0.85$. $J = 1.74$\,meV is assumed.
  }%
\end{figure}

A semiclassical treatment can only explain the line broadening in terms of magnon decay~\cite{Chernyshev2006,Starykh2006,Chernyshev2009,Mourigal2013a,Zhitomirsky2013}. NLSW theory at $H = 0$ describes the spin fluctuations around the 120$^\circ$ ordered state by incorporating single-to-two magnon decay at the leading order $O(S^0)$. At this order, the two-magnon continuum is evaluated by convoluting LSW frequencies. The self-energies include Hartree-Fock decoupling terms as well as the bubble Feynman diagrams comprising a pair of cubic vertices $\Gamma_3 \sim O(S^{1/2})$~\cite{Chernyshev2006,Starykh2006,Chernyshev2009,Mourigal2013a}, with the latter computed with the off-shell treatment. The most crucial one corresponds to the single-to-two magnon decay (see the inset of Fig.~\ref{fig:120deg}a),
\begin{align}
  \Sigma(\mathbf{q},\omega)
  = \frac{1}{2N} \sum_{\mathbf{k}}
  \frac{\left\lvert{\Gamma_3(\mathbf{k}, \mathbf{q} - \mathbf{k}; \mathbf{q})}\right\rvert^2}{\omega - \omega^{H=0}_{\mathbf{k}} - \omega^{H=0}_{\mathbf{q} - \mathbf{k}} + i0},
\end{align}
where $\omega^{H=0}_{\mathbf{k}}$ denotes the zero-field magnon dispersion. We show the zero-field dynamical structure factor, $\mathcal{S}_{H=0}^{\mathrm{tot}}(\mathbf{q},\omega)$, at the M point for representative parameters in Figs.~\ref{fig:120deg}a--\ref{fig:120deg}d. The NLSW result for the ideal TLHAFM ($J_c = 0$ and $\Delta = 1$) exhibits sizable broadening and a strong two-magnon continuum~\cite{Chernyshev2006,Starykh2006,Chernyshev2009,Mourigal2013a} (see also Fig.~\ref{fig:120deg}e). However, a slight deviation from $\Delta = 1$ renders the decay process \emph{ineffective} because the kinematic condition, $\omega^{H=0}_{\mathbf{q}} = \omega^{H=0}_{\mathbf{k}} + \omega^{H=0}_{\mathbf{q} - \mathbf{k}}$, can no longer be fulfilled in 2D for \emph{any} decay vertex over the entire Brillouin zone if $\Delta \lesssim 0.92$~\cite{Chernyshev2006}. This situation can be inferred from the result for $J_c = 0$ and $\Delta = 0.85$, where the two-magnon continuum is pushed to higher frequencies, detached from the single-magnon peaks. In fact, the sharp magnon lines are free from broadening. The suppression of decay results from gapping out one of the two Nambu-Goldstone modes upon lowering the Hamiltonian symmetry from SU(2) to U(1), which greatly reduces the phase space for magnon decay. The interlayer coupling renders the single-magnon peaks even sharper and the continuum even weaker (Figs~\ref{fig:120deg}c and \ref{fig:120deg}d).

To determine whether the anomalous zero-field spin dynamics can be explained by a conventional $1/S$ expansion, it is crucial to estimate $\Delta$ very accurately. Previous experiments reported $\Delta = 0.95$ (low-field electron spin resonance experiments compared with LSW theory~\cite{Susuki2013}) and $\Delta = 0.89$ (zero-field INS experiments compared with NLSW theory~\cite{Ma2016}). However, the NLSW calculation reported a large renormalization of the magnon bandwidth ($\approx 40\%$ reduction relative to the LSW theory)~\cite{Ma2016}, suggesting that the previous estimates of $\Delta$ may be inaccurate. Particularly, given that $\Delta$ is extracted by fitting the induced gap $\propto \! \sqrt{1 - \Delta}$, the LSW approximation underestimates $1 - \Delta$ (deviation from the isotropic exchange) because it \emph{overestimates} the proportionality constant~\cite{Ma2016}.

Figures~\ref{fig:120deg}d and \ref{fig:120deg}f show $\mathcal{S}_{H=0}^{\mathrm{tot}}(\mathbf{q},\omega)$ for \estJc and \estD. We find that $\mathcal{S}_{H=0}^{\mathrm{tot}}(\mathbf{q},\omega)$ remains essentially semiclassical, with sharp magnon lines and a weak continuum, which deviates significantly from the recent results of INS experiments~\cite{Ma2016,Ito2017}. We thus conclude that the Hamiltonian that reproduces the plateau dynamics fails to do so at $H = 0$ \emph{within the spin wave theory}, even after taking magnon-magnon interactions into account at the $1/S$ level. We also mention that the breakdown of the kinematic condition for single-to-two magnon decay also implies the breakdown of the condition for magnon decay into an arbitrary number of magnons~\cite{Zhitomirsky2013}. Thus, the semiclassical picture of weakly interacting magnons is likely inadequate to \emph{simultaneously} explain the low-energy dispersions and the intrinsic incoherent features (such as the high-intensity continuum and the line-broadening) observed in Ba$_3$CoSb$_2$O$_9$ at $H = 0$.

One may wonder if extrinsic effects can explain these experimental observations. It is possible for exchange disorder to produce continuous excitations as in the effective spin-1/2 triangular antiferromagnet  YbMgGaO$_4$ \cite{Paddison2016}. However, our single crystals are the high-quality samples used in Refs.~\cite{Zhou2012,Ma2016}, essentially free from Co$^{2+}$-Sb$^{5+}$ site-disorder. Indeed, our crystals show only one sharp peak at 3.6 K in the zero-field specific heat~\cite{Zhou2012} in contrast to the previous reports of multiple peaks~\cite{Shirata2012}, which may indicate multi-domain structure. Another possible extrinsic effect is the magnon-phonon coupling, that has been invoked to explain the measured spectrum of the spin-3/2 TLHAFM CuCrO$_2$~\cite{Park2016}. However, if that effect were present at zero field, it should also be present in the  UUD state. The fact that Eq.~\eqref{eq:model} reproduces the measured excitation spectrum of the UUD state suggests that the magnon-phonon coupling is negligibly small (a similar line of reasoning can also be applied to the effect of disorder). Indeed, we have also measured the phonon spectrum of Ba$_3$CoSb$_2$O$_9$ in zero field by INS and found no strong signal of magnon-phonon coupling. Our results then suggest that the dynamics of the spin-1/2 TLHAFM is dominated by intrinsic quantum mechanical effects that escape a semiclassical spin-wave description. This situation is analogous to the $(\pi,0)$ wave-vector anomaly observed in various spin-1/2 square-lattice Heisenberg antiferromagnets,~\cite{Christensen2007,Tsyrulin2009,Headings2010,Plumb2014b,Dalla-Piazza2015} but now extending to the entire Brillouin zone in the triangular lattice. Given recent theoretical success on the square-lattice~\cite{Powalski2015}, our results motivate new non-perturbative studies of the spin-1/2 TLHAFM.

\section*{Methods}
\subsection*{Neutron scattering measurements.}
The neutron diffraction data under magnetic fields applied in the [1,-1,0] direction were obtained by using CG-4C cold triple-axis spectrometer with the neutron energy fixed at 5.0\,meV at High Flux Isotope Reactor (HFIR), Oak Ridge National Laboratory (ORNL). The nuclear structure of the crystal was determined at the HB-3A four-circle neutron diffractometer at HFIR, ORNL and then was used to fit the nuclear reflections measured at the CG-4C to confirm that the data reduction is valid. Only the scale factor was refined for fitting the nuclear reflections collected at CG-4C and was also used to scale the moment size for the magnetic structure refinement. 14 magnetic Bragg peaks collected at CG-4C at 10\,T were used for the magnetic structure refinement. The UUD spin configuration with the spins along the field direction was found to best fit the data. The nuclear and magnetic structure refinements were carried out using FullProf Suite~\cite{Rodriguez-Carvajal1993}.

Our inelastic neutron scattering experiments were carried out with the Multi Axis Crystal Spectrometer (MACS)~\cite{MACS} at NIST Center for Neutron Research (NCNR), NIST, and the cold neutron triple-axis spectrometer (V2-FLEXX)~\cite{FLEXX} at Helmholtz-Zentrum Berlin (HZB). The final energies were fixed at 3\,meV and 5\,meV on the MACS and 3.0\,meV on V2-FLEXX.

\subsection*{Constraint on $J$ due to the saturation field.}
An exact expression for the saturation field for $\mathbf{H} \parallel \hat{c}$, $H_\textrm{sat}$, can be obtained from the level crossing condition between the fully polarized state and the ground state in the single-spin-flip sector. From the corresponding expression, we obtain:
\begin{align}
  J = \frac{
    g_\parallel \muB H_\textrm{sat} S^{-1}
  }{
    3 + 6 \Delta + 2 (1 + \Delta) (J_c/J)
  },
\end{align}
where $g_\parallel = 3.87$ and $\mu_0 H_\mathrm{sat} = 32.8$\,T~\cite{Susuki2013}.

\subsection*{Variational analysis on classical instability of the 1/3 plateau in quasi-2D TLHAFMs.}
\begin{figure}[t]
  \centering
  \includegraphics[width=0.65\hsize, bb=0 0 649 244]{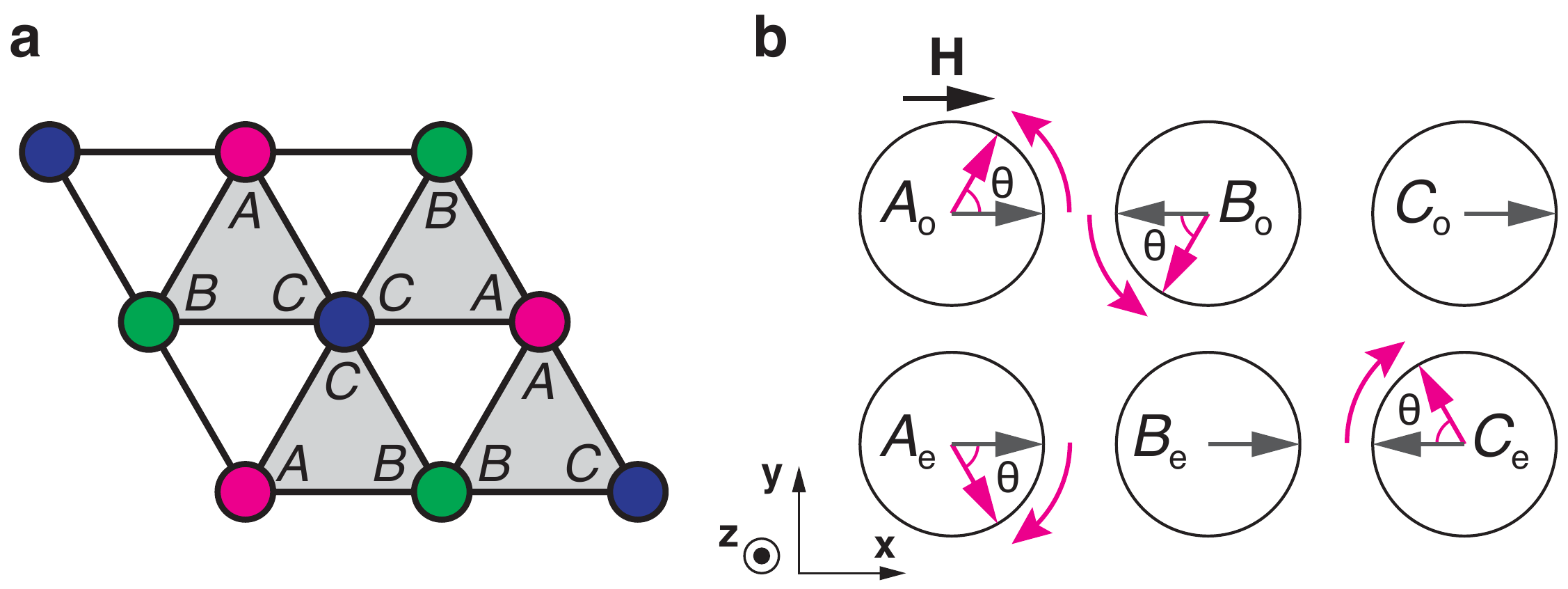}
  \caption{%
    \label{fig:variational}
    \bsf{Classical instability of the UUD state in the quasi-2D lattice.}
    (\bsf{a}) Three-sublattice structure for a single layer and a decomposition of the intralayer bonds into corner-sharing triangles.
    (\bsf{b}) Deformation of the UUD state (see Fig.~\ref{fig:UUD}a) parameterized by $\theta$ shown in the projection in the $ab$ (or $xy$) plane; the spins in sublattices $B_{\mathrm{e}}$ and $C_{\mathrm{o}}$ are unchanged.
  }%
\end{figure}

We show that the UUD state is not the classical ground state in the presence of the antiferromagnetic interlayer exchange $J_c > 0$. To verify that the classical ground space for $J_c = 0$ acquires accidental degeneracy in the in-plane magnetic field, we rewrite Eq.~\eqref{eq:model} as
\begin{align}
  \ham &=
  \frac{J}{2} \sum_{\triangle} \left( \mathbf{S}_{\triangle,A} + \mathbf{S}_{\triangle,B} + \mathbf{S}_{\triangle,C} - \frac{\gmuH}{3J}\hat{\mathbf{x}}\right)^2
  \notag\\
  &\hspace{10pt}
  - \left(1 - \Delta\right) J \sum_{\langle{\mathbf{r}\mathbf{r}'}\rangle} S^z_{\mathbf{r}} S^z_{\mathbf{r}'}
  + J_c \sum_{\mathbf{r}} \left( S^x_{\mathbf{r}} S^x_{\mathbf{r}+\frac{\hat{\mathbf{c}}}{2}} + S^y_{\mathbf{r}} S^y_{\mathbf{r}+\frac{\hat{\mathbf{c}}}{2}} + \Delta S^z_{\mathbf{r}} S^z_{\mathbf{r}+\frac{\hat{\mathbf{c}}}{2}} \right)  + \text{const.},
  \label{eq:H:2}
\end{align}
where the summation of $\sum_{\triangle}$ is taken over the corner-sharing triangles in each layer, with $\mathbf{r} = (\triangle,\mu)$ ($\mu=A,B,C$) denoting the sublattice sites in each triangle. This simply provides an alternative view of each triangular lattice layer (Fig.~\ref{fig:variational}a). $\hat{\mathbf{x}}$ is the unit vector in the $x$ or field direction. The easy-plane anisotropy forces every spin of the classical ground state to lie in the $ab$ plane and the second term in Eq.~\eqref{eq:H:2} has no contribution. Hence, for $J_c = 0$, any three-sublattice spin configuration satisfying $S^z_{\mathbf{r}} = 0$ and
\begin{align}
  \mathbf{S}_{\triangle,A} + \mathbf{S}_{\triangle,B} + \mathbf{S}_{\triangle,C} = \frac{\gmuH}{3J}\hat{\mathbf{x}},~~\forall\triangle,
  \label{eq:degeneracy}
\end{align}
is a classical ground state, where we momentarily regard $\mathbf{S}_{\triangle,\mu}$ as three-component classical spins of length $S$. Since there are only two conditions corresponding to the $x$ and $y$ components of Eq.~\eqref{eq:degeneracy}, whereas three angular variables are needed to specify the three-sublattice state in the $ab$ plane, the classical ground state manifold for $J_c = 0$ retains an accidental degeneracy, similar to the well-known case of the Heisenberg model ($\Delta = 1$)~\cite{Kawamura1985}. The UUD state is the classical ground state only for $\gmuH = 3 J S$.

The classical instability of the UUD state for $J_c > 0$ can be demonstrated by a variational analysis. The UUD state in the 3D lattice enforces frustration for one third of the antiferromagnetic interlayer bonds, inducing large variance of the interlayer interaction. As shown in Fig.~\ref{fig:UUD}a, only two of the three spin pairs along the $\mathbf{c}$ axis per magnetic unit cell can be antiferromagnetically aligned, as favored by $J_c$, while the last one has to be aligned ferromagnetically. To seek for a better classical solution, we consider a deformation of the spin configuration parameterized by $0 \le \theta \le \pi$ at $\gmuH = 3 J S$, such that the spin structure becomes noncollinear within the $ab$ plane (Fig.~\ref{fig:variational}b). Because the magnetization in each layer is fixed at $S/3$ per spin, the sum of the energies associated with the intralayer interaction and the Zeeman coupling is unchanged under this deformation. In the meantime, the energy per magnetic unit cell of the interlayer coupling is varied as
\begin{align}
  E_{c}(\theta) = 2 J_c S^2 \left( \cos2\theta - 2 \cos\theta \right).
\end{align}
We find that $E_{c}(\theta)$ is minimized at $\theta = \pi/3$ for $J_c>0$, corresponding to a saddle point. This is a rather good approximation of the actual classical ground state for small $J_c > 0$, as can be demonstrated by direct minimization of the classical energy obtained from Eq.~\eqref{eq:model}. The crucial observation is that the classical ground state differs from the $\theta = 0$ UUD state.

\subsection*{NLSW calculation for the UUD state.}
We summarize the derivation of the spin wave spectrum in the quasi-2D TLHAFM with easy-plane anisotropy [see Eq.~\eqref{eq:model}]. As discussed in the main text, we first work on the 2D limit $J_c = 0$ exactly at $\gmuH = 3JS$, and a given value of $0 \le \Delta \le 1$, which are the conditions for the UUD state to be the classical ground state. Defining the UUD state as shown in Fig.~\ref{fig:UUD}a, we introduce the Holstein-Primakoff bosons, $a^{(\dag)}_{\mu,\mathbf{r}}$ as in Eqs.~\eqref{eq:rotation:1}--\eqref{eq:HP}. After performing a Fourier transformation,
$a^{\;}_{\mu,\mathbf{k}} = (1/\Ncell)^{1/2} \sum_{\mathbf{r} \in \mu} e^{-i\mathbf{k}\cdot\mathbf{r}} a_{\mu,\mathbf{r}}$,
where $\Ncell = N/6$ is the number of magnetic unit cells (six spins for each) and $N$ is the total number of spins, we obtain the quadratic Hamiltonian as the sum of even layers (sublattices ) and odd layers (sublattices $A_{\mathrm{o}}$--$C_{\mathrm{o}}$) contributions:
\begin{align}
  \Hknot = \HknotE + \HknotO\,,
  \label{eq:H0}
\end{align}
where the constant term has been dropped. Here, 
\begin{align}
  \HknotE
  = \frac{S}{2} \sum_{\mathbf{k} \in \text{RBZ}}
  \begin{pmatrix}
    (\mathbf{a}^{\dagger}_{\mathbf{k}})^\textrm{T} & (\mathbf{a}^{\;}_{-\mathbf{k}})^\textrm{T}
  \end{pmatrix}
  \begin{pmatrix}
    H^0_{11,\mathbf{k}} & H^0_{12,\mathbf{k}} \\[3pt]
    H^0_{21,\mathbf{k}} & H^0_{22,\mathbf{k}}
  \end{pmatrix}
  \begin{pmatrix}
    \mathbf{a}^{\;}_{\mathbf{k}} \\[3pt]
    \mathbf{a}^{\dagger}_{-\mathbf{k}}
  \end{pmatrix},
  \label{eq:H2DLSWeven}
\end{align}
with $H^0_{11,\mathbf{k}} = H^0_{22,\mathbf{k}}$, $H^0_{12,\mathbf{k}} = H^0_{21,\mathbf{k}}$, where the summation over $\mathbf{k}$ is taken in the reduced Brillouin zone (RBZ) corresponding to the magnetic unit cell of the UUD state. From now on, we will denote this summation as $\sum_\mathbf{k}$. We have introduced vector notation for the operators,
\begin{align}
  \mathbf{a}^{}_{\mathbf{k}} 
  =  
  \begin{pmatrix}
    a^{\;}_{A_\mathrm{e},\mathbf{k}} \\[3pt]
    a^{\;}_{B_\mathrm{e},\mathbf{k}} \\[3pt]
    a^{\;}_{C_\mathrm{e},\mathbf{k}}
  \end{pmatrix}
  \equiv
  \begin{pmatrix}
    a^{\;}_{1,\mathbf{k}} \\[3pt]
    a^{\;}_{2,\mathbf{k}} \\[3pt]
    a^{\;}_{3,\mathbf{k}}
  \end{pmatrix},~~
  \mathbf{a}^{\dag}_{-\mathbf{k}} 
  = 
  \begin{pmatrix}
    a^{\dag}_{A_\mathrm{e},-\mathbf{k}} \\[3pt]
    a^{\dag}_{B_\mathrm{e},-\mathbf{k}} \\[3pt]
    a^{\dag}_{C_\mathrm{e},-\mathbf{k}}
  \end{pmatrix}
  \equiv 
  \begin{pmatrix}
    a^{\dag}_{1,-\mathbf{k}} \\[3pt]
    a^{\dag}_{2,-\mathbf{k}} \\[3pt]
    a^{\dag}_{3,-\mathbf{k}}
  \end{pmatrix},
\end{align}
and matrix notation for the quadratic coefficients,
\begin{align}
  H^0_{11,\mathbf{k}}
  &=
  \begin{pmatrix} 
    S^{-1} \gmuH                                  &
    \frac{3}{2}J(1 + \Delta) \gamma^{}_{{\bf k}}  &
    \frac{3}{2}J(1 - \Delta) \gamma^{}_{-{\bf k}} \\[5pt]
    \frac{3}{2}J(1 + \Delta) \gamma^{}_{-{\bf k}} &
    S^{-1} \gmuH                                  &
    \frac{3}{2}J(1 - \Delta) \gamma^{}_{{\bf k}}  \\[5pt]
    \frac{3}{2}J(1 - \Delta) \gamma^{}_{{\bf k}}  &
    \frac{3}{2}J(1 - \Delta) \gamma^{}_{-{\bf k}} &
    6J - S^{-1} \gmuH
  \end{pmatrix},
  \allowdisplaybreaks[1]
  \notag\\[3pt]
  H^0_{12,\mathbf{k}}
  &=
  \begin{pmatrix}
    0                                              &
    -\frac{3}{2}J(1 - \Delta) \gamma^{}_{{\bf k}}  &
    -\frac{3}{2}J(1 + \Delta) \gamma^{}_{-{\bf k}} \\[5pt]
    -\frac{3}{2}J(1 - \Delta) \gamma^{}_{-{\bf k}} &
    0                                              &
    -\frac{3}{2}J(1 + \Delta) \gamma^{}_{{\bf k}}  \\[5pt]
    -\frac{3}{2}J(1 + \Delta) \gamma^{}_{{\bf k}}  &
    -\frac{3}{2}J(1 + \Delta) \gamma^{}_{-{\bf k}} &
    0
  \end{pmatrix},
\end{align}
with $\gamma_\mathbf{k} = \frac{1}{3} (e^{i\mathbf{k}\cdot\mathbf{a}} + e^{i\mathbf{k}\cdot\mathbf{b}} + e^{-i\mathbf{k}\cdot(\mathbf{a} + \mathbf{b})})$.
Similarly, we have
\begin{align}
  \HknotO
  &= \frac{S}{2} \sum_{\mathbf{k}}
  \begin{pmatrix}
    (\bar{\mathbf{a}}^{\dagger}_{\mathbf{k}})^\mathrm{T} & (\bar{\mathbf{a}}^{\;}_{-\mathbf{k}})^\mathrm{T}
  \end{pmatrix}
  \begin{pmatrix}
    \bar{H}^0_{11,\mathbf{k}} & \bar{H}^0_{12,\mathbf{k}} \\[3pt]
    \bar{H}^0_{21,\mathbf{k}} & \bar{H}^0_{22,\mathbf{k}}
  \end{pmatrix}
  \begin{pmatrix}
    \bar{\mathbf{a}}^{\;}_{\mathbf{k}} \\[3pt]
    \bar{\mathbf{a}}^{\dagger}_{-\mathbf{k}}
  \end{pmatrix},
  \label{eq:H2DLSWodd}  
\end{align}
with
\begin{align}
  \bar{\mathbf{a}}^{}_{\mathbf{k}} 
  =  
  \begin{pmatrix}
    a^{\;}_{A_\mathrm{o},\mathbf{k}} \\[3pt]
    a^{\;}_{B_\mathrm{o},\mathbf{k}} \\[3pt]
    a^{\;}_{C_\mathrm{o},\mathbf{k}}
  \end{pmatrix}
  \equiv
  \begin{pmatrix}
    a^{\;}_{4,\mathbf{k}} \\[3pt]
    a^{\;}_{5,\mathbf{k}} \\[3pt]
    a^{\;}_{6,\mathbf{k}}
  \end{pmatrix},~~
  \bar{\mathbf{a}}^{\dag}_{-\mathbf{k}} 
  = 
  \begin{pmatrix}
    a^{\dag}_{A_\mathrm{o},-\mathbf{k}} \\[3pt]
    a^{\dag}_{B_\mathrm{o},-\mathbf{k}} \\[3pt]
    a^{\dag}_{C_\mathrm{o},-\mathbf{k}}
  \end{pmatrix}
  \equiv
  \begin{pmatrix}
    a^{\dag}_{4,-\mathbf{k}} \\[3pt]
    a^{\dag}_{5,-\mathbf{k}} \\[3pt]
    a^{\dag}_{6,-\mathbf{k}}
  \end{pmatrix},
\end{align}
and
\begin{align}
  \bar{H}^0_{11,\mathbf{k}}
  &=
  \bar{H}^0_{22,\mathbf{k}}
  = 
  \begin{pmatrix}
    0 & 1 & 0\\
    0 & 0 & 1\\
    1 & 0 & 0
  \end{pmatrix}
  H^0_{11,\mathbf{k}}
  \begin{pmatrix}
    0 & 0 & 1\\
    1 & 0 & 0\\
    0 & 1 & 0
  \end{pmatrix},
  \notag\\
  \bar{H}^0_{12,\mathbf{k}}
  &=
  \bar{H}^0_{21,\mathbf{k}}
  =
  \begin{pmatrix}
    0 & 1 & 0\\
    0 & 0 & 1\\
    1 & 0 & 0
  \end{pmatrix}
  H^0_{12,\mathbf{k}}
  \begin{pmatrix}
    0 & 0 & 1\\
    1 & 0 & 0\\
    0 & 1 & 0
  \end{pmatrix}.
\end{align}

\begin{figure}[t]
  \centering
  \includegraphics[width=0.9\hsize, bb=0 0 646 236]{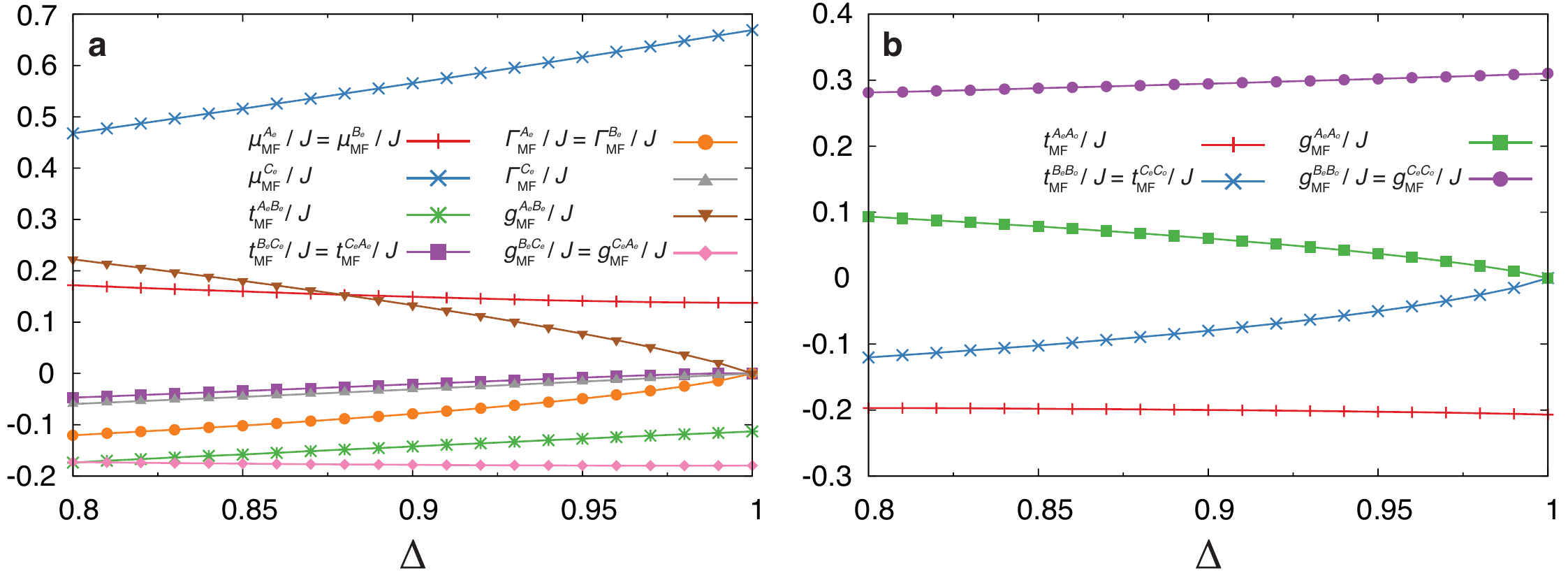}
  \caption{%
    \label{fig:MFs}
    \bsf{$\Delta$-dependence of the recombined MF parameters.}
    MF parameters associated with 
    ({\bf a}) the intra-layer coupling and ({\bf b}) the inter-layer coupling.
  }%
\end{figure}

The excitation spectrum of $\Hknot$ retains two relativistic modes at $\mathbf{k} = 0$ (Fig.~\ref{fig:NLSW}b). Below, we include nonlinear terms to gap out these excitations. At this stage, the nonlinear terms correspond to the mean-field (MF) decoupling of the intra-layer quartic terms. Once we obtain such a MF Hamiltonian with the gapped spectrum, the deviation from the fine-tuned  magnetic field $\gmuH = 3JS$ and  interlayer coupling (as well as some other perturbation, if any) can be included. Here, the additional term contains both LSW and NLSW terms. To proceed, we first define the following mean-fields (MFs) symmetrized by using translational and rotational invariance:
\begin{align}
  \rho^{}_\mu
  &= \frac{1}{\Ncell} \sum_{\mathbf{r} \in \mu} \MF{a^{\dagger}_{\mu,\mathbf{r}} a^{\;}_{\mu,\mathbf{r}}},
  \notag\\
  \delta^{}_\mu
  &= \frac{1}{\Ncell} \sum_{\mathbf{r} \in \mu} \MF{(a^{\;}_{\mu,\mathbf{r}})^2},
  \allowdisplaybreaks[1]
  \notag\\
  \xi^{}_{\mu\nu}
  &= \frac{1}{3\Ncell} \sum_{\mathbf{r} \in \mu} \sum_{\hat{\eta}_{\mu\nu}^{}} \MF{a^{\dagger}_{\mu,\mathbf{r}} a^{\;}_{\nu,\mathbf{r}+\hat{\eta}_{\mu\nu}^{}}},
  \allowdisplaybreaks[1]
  \notag\\
  \zeta^{}_{\mu\nu}
  &= \frac{1}{3\Ncell} \sum_{\mathbf{r} \in \mu} \sum_{\hat{\eta}_{\mu\nu}^{}} \MF{a^{\;}_{\mu,\mathbf{r}} a^{\;}_{\nu,\mathbf{r}+\hat{\eta}_{\mu\nu}^{}}},
  \label{eq:MF}
\end{align}
where $\hat{\eta}_{\mu\nu}^{}$ represents the in-plane displacement vector connecting sites $\mathbf{r} \in \mu$ to a nearest neighbor site in sublattice $\nu$. The mean values $\MF{...}$ are evaluated with the ground state of $\Hknot$. The MFs for odd (even) layers are obtained  from those for even (odd) layers as
\begin{align}
  &
  \rho^{}_{A_\mathrm{o}} = \rho^{}_{B_\mathrm{e}},~~
  \rho^{}_{B_\mathrm{o}} = \rho^{}_{C_\mathrm{e}},~~
  \rho^{}_{C_\mathrm{o}} = \rho^{}_{A_\mathrm{e}},~~
  \allowdisplaybreaks[1]
  \notag\\
  &
  \delta^{}_{A_\mathrm{o}} = \delta^{}_{B_\mathrm{e}},~~
  \delta^{}_{B_\mathrm{o}} = \delta^{}_{C_\mathrm{e}},~~
  \delta^{}_{C_\mathrm{o}} = \delta^{}_{A_\mathrm{e}},~~
  \allowdisplaybreaks[1]
  \notag\\
  &
  \xi^{}_{A_\mathrm{o}B_\mathrm{o}} = \xi^{}_{B_\mathrm{e}C_\mathrm{e}},~~
  \xi^{}_{B_\mathrm{o}C_\mathrm{o}} = \xi^{}_{C_\mathrm{e}A_\mathrm{e}},~~
  \xi^{}_{C_\mathrm{o}A_\mathrm{o}} = \xi^{}_{A_\mathrm{e}B_\mathrm{e}},~~
  \allowdisplaybreaks[1]
  \notag\\
  &
  \zeta^{}_{A_\mathrm{o}B_\mathrm{o}} = \zeta^{}_{B_\mathrm{e}C_\mathrm{e}},~~
  \zeta^{}_{B_\mathrm{o}C_\mathrm{o}} = \zeta^{}_{C_\mathrm{e}A_\mathrm{e}},~~
  \zeta^{}_{C_\mathrm{o}A_\mathrm{o}} = \zeta^{}_{A_\mathrm{e}B_\mathrm{e}}.
\end{align}
By collecting all the contributions mentioned above, we obtain the NLSW Hamiltonian,
  \begin{align}
    \HintSW = \frac{S}{2} \sum_{\mathbf{k}}
    \begin{pmatrix}
      (\mathbf{a}^{\dagger}_{\mathbf{k}})^\mathrm{T}       \hspace{-3pt}&
      (\bar{\mathbf{a}}^{\dagger}_{\mathbf{k}})^\mathrm{T} \hspace{-3pt}&
      (\mathbf{a}^{\;}_{-\mathbf{k}})^\mathrm{T}           \hspace{-3pt}&
      (\bar{\mathbf{a}}^{\;}_{-\mathbf{k}})^\mathrm{T}     \hspace{-0pt}
    \end{pmatrix}
    \begin{pmatrix}
      \matHee{\mathbf{k}}^{} &
      \matHeo{\mathbf{k}}^{} &
      \matHee{\mathbf{k}}'   &
      \matHeo{\mathbf{k}}'
      \\[5pt]
      \bigl(\matHeo{\mathbf{k}}^{}\bigr)_{}^{\dag}   &
      \matHoo{\mathbf{k}}^{} &
      \bigl(\matHeo{-\mathbf{k}}'\bigr)_{}^{\mathrm{T}} &
      \matHoo{\mathbf{k}}'
      \\[5pt]
      \bigl(\matHee{-\mathbf{k}}'\bigr)_{}^{\ast}   &
      \bigl(\matHeo{-\mathbf{k}}'\bigr)_{}^{\ast}   &
      \bigl(\matHee{-\mathbf{k}}^{}\bigr)_{}^{\ast} &
      \bigl(\matHeo{-\mathbf{k}}^{}\bigr)_{}^{\ast}
      \\[5pt]
      \bigl(\matHeo{\mathbf{k}}'\bigr)_{}^{\dag}    &
      \bigl(\matHoo{-\mathbf{k}}'\bigr)_{}^{\ast}   &
      \bigl(\matHeo{-\mathbf{k}}^{}\bigr)_{}^{\mathrm{T}} &
      \bigl(\matHoo{-\mathbf{k}}^{}\bigr)_{}^{\ast}
    \end{pmatrix}
    \begin{pmatrix}
      \mathbf{a}^{\;}_{\mathbf{k}} \\[5pt]
      \bar{\mathbf{a}}^{\;}_{\mathbf{k}} \\[5pt]
      \mathbf{a}^{\dagger}_{-\mathbf{k}} \\[5pt]
      \bar{\mathbf{a}}^{\dagger}_{-\mathbf{k}}
    \end{pmatrix},
    \label{eq:H-NLSW}
  \end{align}
  where
  \begin{align}
    \matHee{\mathbf{k}}^{}
    &
    = H^0_{11,\mathbf{k}}
    +
    \begin{pmatrix} 
      -2 J_c & 0      & 0      \\
      0      & 2 J_c  & 0      \\
      0      & 0      & 2 J_c
    \end{pmatrix}
    + S^{-1}
    \begin{pmatrix}
      \mu^{A_\mathrm{e}}_\textrm{MF} + 2 J_c \rho^{}_{A_\mathrm{o}} &
      \left(t^{A_\mathrm{e}B_\mathrm{e}}_\textrm{MF}\right)^{\ast} \gamma^{}_{\mathbf{k}} &
      t^{C_\mathrm{e}A_\mathrm{e}}_\textrm{MF} \gamma^{}_{-\mathbf{k}}
      \\[5pt]
      t^{A_\mathrm{e}B_\mathrm{e}}_\textrm{MF} \gamma^{}_{-\mathbf{k}} &
      \mu^{B_\mathrm{e}}_\textrm{MF} -2 J_c \rho^{}_{B_\mathrm{o}} &
      \left(t^{B_\mathrm{e}C_\mathrm{e}}_\textrm{MF}\right)^{\ast} \gamma^{}_{\mathbf{k}} 
      \\[5pt]
      \left(t^{C_\mathrm{e}A_\mathrm{e}}_\textrm{MF}\right)^{\ast} \gamma^{}_{\mathbf{k}} &
      t^{B_\mathrm{e}C_\mathrm{e}}_\textrm{MF} \gamma^{}_{-\mathbf{k}} &
      \mu^{C_\mathrm{e}}_\textrm{MF} - 2 J_c \rho^{}_{C_\mathrm{o}}
    \end{pmatrix},
    \allowdisplaybreaks[1]
    \notag\\[3pt]
    \matHoo{\mathbf{k}}^{}
    &
    = \bar{H}^0_{11,\mathbf{k}}
    +
    \begin{pmatrix} 
      -2 J_c & 0      & 0      \\
      0      & 2 J_c  & 0      \\
      0      & 0      & 2 J_c
    \end{pmatrix}
    + S^{-1}
    \begin{pmatrix}
      \mu^{A_\mathrm{o}}_\textrm{MF} + 2 J_c \rho^{}_{A_\mathrm{e}} &
      \left(t^{A_\mathrm{o}B_\mathrm{o}}_\textrm{MF}\right)^{\ast} \gamma^{}_{\mathbf{k}} &
      t^{C_\mathrm{o}A_\mathrm{o}}_\textrm{MF} \gamma^{}_{-\mathbf{k}}
      \\[5pt]
      t^{A_\mathrm{o}B_\mathrm{o}}_\textrm{MF} \gamma^{}_{-\mathbf{k}} &
      \mu^{B_\mathrm{o}}_\textrm{MF} - 2 J_c \rho^{}_{B_\mathrm{e}} &
      \left(t^{B_\mathrm{o}C_\mathrm{o}}_\textrm{MF}\right)^{\ast} \gamma^{}_{\mathbf{k}} 
      \\[5pt]
      \left(t^{C_\mathrm{o}A_\mathrm{o}}_\textrm{MF}\right)^{\ast} \gamma^{}_{\mathbf{k}} &
      t^{B_\mathrm{o}C_\mathrm{o}}_\textrm{MF} \gamma^{}_{-\mathbf{k}} &
      \mu^{C_\mathrm{o}}_\textrm{MF} - 2 J_c \rho^{}_{C_\mathrm{e}}
    \end{pmatrix},
    \allowdisplaybreaks[1]
    \notag\\[3pt]
    \matHee{\mathbf{k}}'
    &=
    H^0_{12,\mathbf{k}}
    + S^{-1}
    \begin{pmatrix}
      \Gamma^{A_\mathrm{e}}_\textrm{MF} &
      g^{A_\mathrm{e}B_\mathrm{e}}_\textrm{MF} \gamma^{}_{\mathbf{k}}  &
      g^{C_\mathrm{e}A_\mathrm{e}}_\textrm{MF} \gamma^{}_{-\mathbf{k}} \\[4pt]
      g^{A_\mathrm{e}B_\mathrm{e}}_\textrm{MF} \gamma^{}_{-\mathbf{k}} &
      \Gamma^{B_\mathrm{e}}_\textrm{MF} &
      g^{B_\mathrm{e}C_\mathrm{e}}_\textrm{MF} \gamma^{}_{\mathbf{k}} \\[4pt]
      g^{C_\mathrm{e}A_\mathrm{e}}_\textrm{MF} \gamma^{}_{\mathbf{k}}  &
      g^{B_\mathrm{e}C_\mathrm{e}}_\textrm{MF} \gamma^{}_{-\mathbf{k}} &
      \Gamma^{C}_\textrm{MF}
    \end{pmatrix},
    \allowdisplaybreaks[1]
    \notag\\[3pt]
    \matHoo{\mathbf{k}}'
    &=
    \bar{H}^0_{12,\mathbf{k}}
    + S^{-1}
    \begin{pmatrix}
      \Gamma^{A_\mathrm{o}}_\textrm{MF} &
      g^{A_\mathrm{o}B_\mathrm{o}}_\textrm{MF} \gamma^{}_{\mathbf{k}}  &
      g^{C_\mathrm{o}A_\mathrm{o}}_\textrm{MF} \gamma^{}_{-\mathbf{k}} \\[5pt]
      g^{A_\mathrm{o}B_\mathrm{o}}_\textrm{MF} \gamma^{}_{-\mathbf{k}} &
      \Gamma^{B_\mathrm{o}}_\textrm{MF} &
      g^{B_\mathrm{o}C_\mathrm{o}}_\textrm{MF} \gamma^{}_{\mathbf{k}} \\[5pt]
      g^{C_\mathrm{o}A_\mathrm{o}}_\textrm{MF} \gamma^{}_{\mathbf{k}}  &
      g^{B_\mathrm{o}C_\mathrm{o}}_\textrm{MF} \gamma^{}_{-\mathbf{k}} &
      \Gamma^{C_\mathrm{o}}_\textrm{MF}
    \end{pmatrix},
    \allowdisplaybreaks[1]
    \notag\\[3pt]
    \matHeo{\mathbf{k}}^{}
    &=
    \cos k_3
    \begin{pmatrix} 
      J_c \left(1 + \Delta\right) &  0          &   0      \\
      0          &  J_c \left(1 - \Delta\right) &   0      \\
      0          &  0          &  J_c\left(1 - \Delta\right)
    \end{pmatrix}
    + S^{-1} \cos k_3 
    \begin{pmatrix} 
      \left(t^{A_\mathrm{e}A_\mathrm{o}}_{\textrm{MF}}\right)^{\ast} &
      0 & 0
      \\
      0 &
      \left(t^{B_\mathrm{e}B_\mathrm{o}}_{\textrm{MF}}\right)^{\ast} &
      0
      \\
      0 & 0 &
      \left(t^{C_\mathrm{e}C_\mathrm{o}}_{\textrm{MF}}\right)^{\ast} 
    \end{pmatrix},
    \allowdisplaybreaks[1]
    \notag\\[3pt]
    \matHeo{\mathbf{k}}' 
    &= 
    \cos k_3
    \begin{pmatrix} 
      -J_c \left(1 - \Delta\right) &  0          &   0      \\
      0          &  -J_c \left(1 + \Delta\right) &   0      \\
      0          &  0          &  -J_c\left(1 + \Delta\right)
    \end{pmatrix}
    + S^{-1} \cos k_3
    \begin{pmatrix} 
      g_\textrm{MF}^{A_\mathrm{e}A_\mathrm{o}} &  0  &  0  \\
      0  & g_\textrm{MF}^{B_\mathrm{e}B_\mathrm{o}} &  0  \\
      0  & 0  & g_\textrm{MF}^{C_\mathrm{e}C_\mathrm{o}}
    \end{pmatrix}.
  \end{align}
  Here the MF parameters are given as follows. First, those associated with the intralayer coupling are
  \begin{align}
    \mu^{A_\mathrm{e}}_\textrm{MF}
    &= 3J \left[
      {\rho}^{}_{B_\mathrm{e}} - {\rho}^{}_{C_\mathrm{e}}
      - \frac{1 + \Delta}{2} \left( \Re{{\xi}^{}_{A_\mathrm{e}B_\mathrm{e}}} - \Re{{\zeta}^{}_{C_\mathrm{e}A_\mathrm{e}}} \right)
      - \frac{1 - \Delta}{2} \left( \Re{{\xi}^{}_{C_\mathrm{e}A_\mathrm{e}}} - \Re{{\zeta}^{}_{A_\mathrm{e}B_\mathrm{e}}} \right)
      \right],
    \allowdisplaybreaks[1]
    \notag\\
    \mu^{B_\mathrm{e}}_\textrm{MF}
    &= 3J \left[
      {\rho}^{}_{A_\mathrm{e}} - {\rho}^{}_{C_\mathrm{e}}
      - \frac{1 + \Delta}{2} \left( \Re{{\xi}^{}_{A_\mathrm{e}B_\mathrm{e}}} - \Re{{\zeta}^{}_{B_\mathrm{e}C_\mathrm{e}}} \right)
      - \frac{1 - \Delta}{2} \left( \Re{{\xi}^{}_{B_\mathrm{e}C_\mathrm{e}}} - \Re{{\zeta}^{}_{A_\mathrm{e}B_\mathrm{e}}} \right)
      \right],
    \allowdisplaybreaks[1]
    \notag\\
    \mu^{C_\mathrm{e}}_\textrm{MF}
    &= 3J \left[
      -{\rho}^{}_{A_\mathrm{e}} - {\rho}^{}_{B_\mathrm{e}} 
      + \frac{1 + \Delta}{2} \left( \Re{{\zeta}^{}_{B_\mathrm{e}C_\mathrm{e}}} + \Re{{\zeta}^{}_{C_\mathrm{e}A_\mathrm{e}}} \right)
      - \frac{1 - \Delta}{2} \left( \Re{{\xi}^{}_{B_\mathrm{e}C_\mathrm{e}}} + \Re{{\xi}^{}_{C_\mathrm{e}A_\mathrm{e}}} \right)
      \right],
    \allowdisplaybreaks[1]
    \notag\\
    t^{A_\mathrm{e}B_\mathrm{e}}_\textrm{MF} &= 3J \left[ {\xi}^{}_{A_\mathrm{e}B_\mathrm{e}}
      - \frac{1 + \Delta}{4} \left( {\rho}^{}_{A_\mathrm{e}} + {\rho}^{}_{B_\mathrm{e}} \right)
      + \frac{1 - \Delta}{8} \left( {\delta}^{\ast}_{A_\mathrm{e}} + {\delta}^{}_{B_\mathrm{e}} \right) \right],
    \allowdisplaybreaks[1]
    \notag\\
    t^{B_\mathrm{e}C_\mathrm{e}}_\textrm{MF} &= 3J \left[ -{\xi}^{}_{B_\mathrm{e}C_\mathrm{e}}
      + \frac{1 + \Delta}{8} \left( {\delta}^{\ast}_{B_\mathrm{e}} + {\delta}^{}_{C_\mathrm{e}} \right)
      - \frac{1 - \Delta}{4} \left( {\rho}^{}_{B_\mathrm{e}} + {\rho}^{}_{C_\mathrm{e}} \right) \right],
    \allowdisplaybreaks[1]
    \notag\\
    t^{C_\mathrm{e}A_\mathrm{e}}_\textrm{MF} &= 3J \left[ -{\xi}^{}_{C_\mathrm{e}A_\mathrm{e}}
      + \frac{1 + \Delta}{8} \left( {\delta}^{\ast}_{C_\mathrm{e}} + {\delta}^{}_{A_\mathrm{e}} \right)
      - \frac{1 - \Delta}{4} \left( {\rho}^{}_{C_\mathrm{e}} + {\rho}^{}_{A_\mathrm{e}} \right) \right],
    \allowdisplaybreaks[1]
    \notag\\
    \Gamma^{A_\mathrm{e}}_\textrm{MF}
    &= \frac{3J}{2} \left[
      \frac{1 + \Delta}{2}\left(\xi^{}_{C_\mathrm{e}A_\mathrm{e}} - \zeta^{}_{A_\mathrm{e}B_\mathrm{e}}\right)
      + \frac{1 - \Delta}{2}\left(\xi^{\ast}_{A_\mathrm{e}B_\mathrm{e}} - \zeta^{}_{C_\mathrm{e}A_\mathrm{e}}\right)
      \right],
    \allowdisplaybreaks[1]
    \notag\\
    \Gamma^{B_\mathrm{e}}_\textrm{MF}
    &= \frac{3J}{2} \left[
      \frac{1 + \Delta}{2}\left(\xi^{\ast}_{B_\mathrm{e}C_\mathrm{e}} - \zeta^{}_{A_\mathrm{e}B_\mathrm{e}}\right)
      + \frac{1 - \Delta}{2}\left(\xi^{}_{A_\mathrm{e}B_\mathrm{e}} - \zeta^{}_{B_\mathrm{e}C_\mathrm{e}}\right)
      \right],
    \allowdisplaybreaks[1]
    \notag\\
    \Gamma^{C_\mathrm{e}}_\textrm{MF}
    &= \frac{3J}{2} \left[
      \frac{1 + \Delta}{2}\left(\xi^{}_{B_\mathrm{e}C_\mathrm{e}} - \xi^{\ast}_{C_\mathrm{e}A_\mathrm{e}}\right)
      - \frac{1 - \Delta}{2}\left(\zeta^{}_{B_\mathrm{e}C_\mathrm{e}} + \zeta^{}_{C_\mathrm{e}A_\mathrm{e}}\right)
      \right],
    \allowdisplaybreaks[1]
    \notag\\
    g^{A_\mathrm{e}B_\mathrm{e}}_\textrm{MF} &= 3J \left[ {\zeta}^{}_{A_\mathrm{e}B_\mathrm{e}}
      - \frac{1 + \Delta}{8} \left( {\delta}^{}_{A_\mathrm{e}} + {\delta}^{}_{B_\mathrm{e}} \right)
      + \frac{1 - \Delta}{4} \left( {\rho}^{}_{A_\mathrm{e}} + {\rho}^{}_{B_\mathrm{e}} \right) \right],
    \allowdisplaybreaks[1]
    \notag\\
    g^{B_\mathrm{e}C_\mathrm{e}}_\textrm{MF} &= 3J \left[ -{\zeta}^{}_{B_\mathrm{e}C_\mathrm{e}}
      + \frac{1 + \Delta}{4} \left( {\rho}^{}_{B} + {\rho}^{}_{C} \right)
      - \frac{1 - \Delta}{8} \left( {\delta}^{}_{B} + {\delta}^{}_{C} \right)
      \right],
    \allowdisplaybreaks[1]
    \notag\\
    g^{C_\mathrm{e}A_\mathrm{e}}_\textrm{MF} &= 3J \left[ -{\zeta}^{}_{C_\mathrm{e}A_\mathrm{e}}
      + \frac{1 + \Delta}{4} \left( {\rho}^{}_{C_\mathrm{e}} + {\rho}^{}_{A_\mathrm{e}} \right)
      - \frac{1 - \Delta}{8} \left( {\delta}^{}_{C_\mathrm{e}} + {\delta}^{}_{A_\mathrm{e}} \right)
      \right],
  \end{align}
for even layers and
\begin{align}
  &
  \mu_\textrm{MF}^{A_\mathrm{o}} = \mu_\textrm{MF}^{B_\mathrm{e}},~~
  \mu_\textrm{MF}^{B_\mathrm{o}} = \mu_\textrm{MF}^{C_\mathrm{e}},~~
  \mu_\textrm{MF}^{C_\mathrm{o}} = \mu_\textrm{MF}^{A_\mathrm{e}},~~
  \notag\\[2pt]
  &
  t_{\textrm{MF}}^{A_\mathrm{o}B_\mathrm{o}} = t_{\textrm{MF}}^{B_\mathrm{e}C_\mathrm{e}},~~
  t_{\textrm{MF}}^{B_\mathrm{o}C_\mathrm{o}} = t_{\textrm{MF}}^{C_\mathrm{e}A_\mathrm{e}},~~
  t_{\textrm{MF}}^{C_\mathrm{o}A_\mathrm{o}} = t_{\textrm{MF}}^{A_\mathrm{e}B_\mathrm{e}},
  \notag\\[2pt]
  &
  \Gamma_\textrm{MF}^{A_\mathrm{o}} = \Gamma_\textrm{MF}^{B_\mathrm{e}},~~
  \Gamma_\textrm{MF}^{B_\mathrm{o}} = \Gamma_\textrm{MF}^{C_\mathrm{e}},~~
  \Gamma_\textrm{MF}^{C_\mathrm{o}} = \Gamma_\textrm{MF}^{A_\mathrm{e}},
  \allowdisplaybreaks[1]
  \notag\\[2pt]
  &
  g_{\textrm{MF}}^{A_\mathrm{o}B_\mathrm{o}} = g_{\textrm{MF}}^{B_\mathrm{e}C_\mathrm{e}},~~
  g_{\textrm{MF}}^{B_\mathrm{o}C_\mathrm{o}} = g_{\textrm{MF}}^{C_\mathrm{e}A_\mathrm{e}},~~
  g_{\textrm{MF}}^{C_\mathrm{o}A_\mathrm{o}} = g_{\textrm{MF}}^{A_\mathrm{e}B_\mathrm{e}},
\end{align}
for odd layers. Similarly, the new MF parameters associated with the interlayer coupling are
\begin{align}
  t^{A_\mathrm{e}A_\mathrm{o}}_{\textrm{MF}}
  &= J_c \left[
    - \frac{1 + \Delta}{2} \left( \rho^{}_{A_\mathrm{e}} + \rho^{}_{A_\mathrm{o}} \right)
    + \frac{1 - \Delta}{4} \left( \delta^{\ast}_{A_\mathrm{e}} + \delta^{}_{A_\mathrm{o}} \right)
    \right],
  \notag\\[1.5pt]
  t^{B_\mathrm{e}B_\mathrm{o}}_{\textrm{MF}}
  &= J_c \left[
    \frac{1 + \Delta}{4} \left( \delta^{\ast}_{B_\mathrm{e}} + \delta^{}_{B_\mathrm{o}} \right)
    - \frac{1 - \Delta}{2} \left( \rho^{}_{B_\mathrm{e}} + \rho^{}_{B_\mathrm{o}} \right)
    \right],
  \notag\\[1.5pt]
  t^{C_\mathrm{e}C_\mathrm{o}}_{\textrm{MF}}
  &= J_c \left[
    \frac{1 + \Delta}{4} \left( \delta^{\ast}_{C_\mathrm{e}} + \delta^{}_{C_\mathrm{o}} \right)
    - \frac{1 - \Delta}{2} \left( \rho^{}_{C_\mathrm{e}} + \rho^{}_{C_\mathrm{o}} \right)
    \right],
  \notag\\[1.5pt]
  g_\textrm{MF}^{A_\mathrm{e}A_\mathrm{o}}
  &=
  J_c \left[
    - \frac{1 + \Delta}{4} \left( \delta^{}_{A_\mathrm{e}} + \delta^{}_{A_\mathrm{o}} \right)
    + \frac{1 - \Delta}{2} \left( \rho^{}_{A_\mathrm{e}} + \rho^{}_{A_\mathrm{o}} \right)
    \right],
  \allowdisplaybreaks[1]    
  \notag\\[1.5pt]
  g_\textrm{MF}^{B_\mathrm{e}B_\mathrm{o}}
  &=
  J_c \left[
    \frac{1 + \Delta}{2} \left( \rho^{}_{B_\mathrm{e}} + \rho^{}_{B_\mathrm{o}} \right)
    - \frac{1 - \Delta}{4} \left( \delta^{}_{B_\mathrm{e}} + \delta^{}_{B_\mathrm{o}} \right)
    \right],
  \notag\\[1.5pt]
  g_\textrm{MF}^{C_\mathrm{e}C_\mathrm{o}}
  &=
  J_c \left[
    \frac{1 + \Delta}{2} \left( \rho^{}_{C_\mathrm{e}} + \rho^{}_{C_\mathrm{o}} \right)
    - \frac{1 - \Delta}{4} \left( \delta^{}_{C_\mathrm{e}} + \delta^{}_{C_\mathrm{o}} \right)
    \right].
\end{align}
Figure~\ref{fig:MFs} shows the $\Delta$-dependence of these MF parameters.
Because these MF parameters are real valued, the coefficient matrix in Eq.~\eqref{eq:H-NLSW} has the form,
\begin{align}
  H_\textrm{NLSW} =
  \begin{pmatrix}
    P^{}_\mathbf{k} & Q^{}_\mathbf{k} \\
    Q^{}_\mathbf{k} & P^{}_\mathbf{k}
  \end{pmatrix},~~~
\end{align}
with
\begin{align}
  P^{}_\mathbf{k}
  =
  \begin{pmatrix}
    \matHee{\mathbf{k}}^{} &
    \matHeo{\mathbf{k}}^{}
    \\[5pt]
    \matHeo{\mathbf{k}}^{} &
    \matHoo{\mathbf{k}}^{}
  \end{pmatrix},
  ~~~~
  Q^{}_\mathbf{k}
  =
  \begin{pmatrix}
    \matHee{\mathbf{k}}'   &
    \matHeo{\mathbf{k}}'
    \\[5pt]
    \matHeo{\mathbf{k}}'   &
    \matHoo{\mathbf{k}}'
  \end{pmatrix}.
\end{align}
This form can be diagonalized by a Bogoliubov transformation,
\begin{align}
  \begin{pmatrix}
    \mathbf{a}^{\;}_{\mathbf{k}} \\[3pt]
    \bar{\mathbf{a}}^{\;}_{\mathbf{k}} \\[3pt]
    \mathbf{a}^{\dagger}_{-\mathbf{k}} \\[3pt]
    \bar{\mathbf{a}}^{\dagger}_{-\mathbf{k}}
  \end{pmatrix}
  =
  \left(\begin{array}{cccccc}
    &&&&&\\[3pt]
    \multicolumn{3}{c}{\smash{\raisebox{.5\normalbaselineskip}{$U_{\mathbf{k}}$}}} &
    \multicolumn{3}{c}{\smash{\raisebox{.5\normalbaselineskip}{$V_{\mathbf{k}}$}}} \\[3pt]
    &&&&&\\[3pt]
    \multicolumn{3}{c}{\smash{\raisebox{.5\normalbaselineskip}{$V_{\mathbf{k}}$}}} &
    \multicolumn{3}{c}{\smash{\raisebox{.5\normalbaselineskip}{$U_{\mathbf{k}}$}}}
  \end{array}\right)
  \,
  \left(\begin{array}{c}
    \\[3pt]
    \hspace{-5pt}\smash{\raisebox{.5\normalbaselineskip}{$\bm{\alpha}^{\;}_{\mathbf{k}}$}}\hspace{-5pt} \\[3pt]
    \\[3pt]
    \hspace{-5pt}\smash{\raisebox{.5\normalbaselineskip}{$\bm{\alpha}^{\dag}_{-\mathbf{k}}$}}\hspace{-5pt}
  \end{array}\right),
\end{align}
where $\bm{\alpha}^{\;}_{\mathbf{k}}$ ($\bm{\alpha}^{\dag}_{-\mathbf{k}}$) is the 6-component vector comprising the annihilation (creation) operators of Bogoliubov bosons. The transformation matrices satisfy $U_{\mathbf{k}}^{\mu\kappa} = \bigl(U_{-\mathbf{k}}^{\mu\kappa}\bigr)^\ast$ and $V_{\mathbf{k}}^{\mu\kappa} = \bigl(V_{-\mathbf{k}}^{\mu\kappa}\bigr)^\ast$. The poles, $\omega^{}_{\kappa,\mathbf{k}}$, are the square-roots of the eigenvalues of $S^2(P_\mathbf{k} + Q_\mathbf{k}) (P_\mathbf{k} - Q_\mathbf{k})$.

When calculating the sublattice magnetization, the reduction of the ordered moment relative to the classical value $S$ corresponds to the local magnon density. With the phase factors for each sublattice, $c^{}_{A_\mathrm{e}} = c^{}_{B_\mathrm{e}} = -c^{}_{C_\mathrm{e}} = c^{}_{A_\mathrm{o}} = -c^{}_{B_\mathrm{o}} = c^{}_{C_\mathrm{o}} = 1$ (see Fig.~{\ref{fig:UUD}}a), we have
\begin{align}
  \bigl\langle{S^x_\mathbf{r}}\bigr\rangle
  &= c^{}_\mu \left( S - \bigl\langle{ a^{\dag}_{\mu,\mathbf{r}} a^{\;}_{\mu,\mathbf{r}} }\bigr\rangle\right)
  = c^{}_\mu \left( S - \frac{1}{\Ncell} \sum_\mathbf{k} \sum_\kappa \left\lvert{ V_\mathbf{k}^{\mu\kappa} }\right\rvert^2 \right),
\end{align}
for site $\mathbf{r}$ in sublattice $\mu$.

The dynamical spin structure factor is defined by
\begin{align}
  \mathcal{S}^{\alpha\alpha}(\mathbf{q},\omega)
  &= \int_{-\infty}^{\infty} \frac{dt}{2\pi} e^{i\omega t} \frac{1}{N} \sum_{\mathbf{r},\mathbf{r}'} e^{-i\mathbf{q} \cdot (\mathbf{r} - \mathbf{r}')}
  \left\langle{
    S_{\mathbf{r}}^{\alpha}(t) S_{\mathbf{r}'}^{\alpha}(0)
  }\right\rangle
  \notag\\
  &=
  \sum_n \delta(\omega - \omega_n)\,
  \left\lvert{\bra{0} S_{\mathbf{q}}^{\alpha} \ket{n}}\right\rvert^2,
\end{align}
where
$S_{\mathbf{q}}^{\alpha} = N^{-1/2} \sum_\mathbf{r} S_\mathbf{r}^{\alpha} e^{-i\mathbf{q} \cdot \mathbf{r}}$
and
$\ket{n}$ and $\omega_n$ denote the $n$th excited state and its excitation energy, respectively. The longitudinal spin component is
\begin{align}
  S_{\mathbf{q}}^{x}
  &=
  \frac{\sqrt{N}}{3}S \left(\delta_{\mathbf{q},0} + \delta_{\mathbf{q},\mathbf{Q}} + \delta_{\mathbf{q},-\mathbf{Q}}\right)
  + \delta S_{\mathbf{q}}^{x},
\end{align}
with $\mathbf{Q} = (1/3,1/3,1)$ and
\begin{align}
  \delta S_{\mathbf{q}}^{x}
  = - \sqrt{\frac{1}{N}} \sum_{\mu,\mathbf{k}} c^{}_\mu a^{\dag}_{\mu,\mathbf{k} - \mathbf{q}} a^{\;}_{\mu,\mathbf{k}},
\end{align}
We truncate the expansions of the transverse spin components at the lowest order:
\begin{align}
  S_{\mathbf{q}}^{y}
  & \approx
  -i \sqrt{\frac{S}{12}}
  \sum_{\mu}
  \left(
  a^{\;}_{\mu,\mathbf{q}} - a^{\dag}_{\mu,-\mathbf{q}}
  \right),
  \notag\\
  S_{\mathbf{q}}^{z}
  & \approx
  \sqrt{\frac{S}{12}}
  \sum_{\mu} (-c^{}_\mu)
  \left(
  a^{\;}_{\mu,\mathbf{q}} + a^{\dag}_{\mu,-\mathbf{q}}
  \right).
\end{align}
The transverse components of the dynamical structure factor, $\mathcal{S}_{\perp}(\mathbf{q},\omega) = \mathcal{S}^{yy}(\mathbf{q},\omega) + \mathcal{S}^{zz}(\mathbf{q},\omega)$, reveal the magnon dispersion,
\begin{align}
  \mathcal{S}^{yy}(\mathbf{q},\omega)
  &=
  \sum_n \delta(\omega - \omega_n)\,
  \left\lvert{\bra{0} S_{\mathbf{q}}^{y} \ket{n}}\right\rvert^2
  \notag\\
  & \approx
  \frac{S}{12}
  \sum_{\kappa}
  \delta(\omega - \omega^{}_{\kappa,\mathbf{q}})\,
  \Biggl\lvert{ 
    \sum_{\mu}
    \left(U_{\mathbf{q}}^{\mu\kappa} - V_{\mathbf{q}}^{\mu\kappa}\right)
  }\Biggr\rvert^2,
  \notag\\
  \mathcal{S}^{zz}(\mathbf{q},\omega)
  &=
  \sum_n \delta(\omega - \omega_n)\,
  \left\lvert{\bra{0} S_{\mathbf{q}}^{z} \ket{n}}\right\rvert^2
  \notag\\
  & \approx
  \frac{S}{12}
  \sum_{\kappa}
  \delta(\omega - \omega^{}_{\kappa,\mathbf{q}})\,
  \Biggl\lvert{ 
    \sum_{\mu} c^{}_\mu
    \left(U_{\mathbf{q}}^{\mu\kappa} + V_{\mathbf{q}}^{\mu\kappa}\right)
  }\Biggr\rvert^2.
\end{align}
Meanwhile, $\mathcal{S}^{xx}(\mathbf{q},\omega)$ comprises the elastic contribution and the longitudinal fluctuations,
\begin{align}
  \mathcal{S}_{\parallel}(\mathbf{q},\omega)
  &=
  \sum_n \delta(\omega - \omega_n)\,
  \left\lvert{\bra{0} \delta S_{\mathbf{q}}^{x} \ket{n}}\right\rvert^2,
\end{align}
which can be evaluated by using Wick's theorem.
The result at $T = 0$ is
\begin{align}
  \mathcal{S}_{\parallel}(\mathbf{q},\omega)
  &= \Theta(\omega) 
  N^{-1} \sum_{\mathbf{k}}
  \sum_{\kappa,\lambda,\mu,\nu}  
  c^{}_{\mu} c^{}_{\nu}\Re{A}_{\mu\nu;\kappa\lambda}(\mathbf{k};\mathbf{q})
  \,
  \delta(\omega - \omega^{}_{\kappa,-\mathbf{k}+\mathbf{q}} - \omega^{}_{\lambda,\mathbf{k}}),
\end{align}
where
\begin{align}
  A_{\mu\nu;\kappa\lambda}(\mathbf{k};\mathbf{q})
  &= 
  \frac{1}{2}
  \left[
    \left( U_{\mathbf{k}-\mathbf{q}}^{\mu\kappa}\right)^\ast V_{\mathbf{k}}^{\mu\lambda}
    + \left( V_{\mathbf{k}-\mathbf{q}}^{\mu\kappa}\right)^\ast U_{\mathbf{k}}^{\mu\lambda}
    \right]
  \,
  \left[
    U_{\mathbf{k}-\mathbf{q}}^{\nu\kappa} \left( V_{\mathbf{k}}^{\nu\lambda} \right)^\ast
    + V_{\mathbf{k}-\mathbf{q}}^{\nu\kappa} \left(U_{\mathbf{k}}^{\nu\lambda}\right)^\ast
    \right].
\end{align}

\subsection*{Data Availability.}
All relevant data are available from the corresponding authors upon reasonable request.

\bibliographystyle{naturemag_noURL}
\bibliography{ref}

\begin{thebibliography}{10}
\expandafter\ifx\csname url\endcsname\relax
  \def\url#1{\texttt{#1}}\fi
\expandafter\ifx\csname urlprefix\endcsname\relax\def\urlprefix{URL }\fi
\providecommand{\bibinfo}[2]{#2}
\providecommand{\eprint}[2][]{\url{#2}}

\bibitem{Villain1980}
\bibinfo{author}{{Villain, J.}}, \bibinfo{author}{{Bidaux, R.}},
  \bibinfo{author}{{Carton, J.-P.}} \& \bibinfo{author}{{Conte, R.}}
\newblock \bibinfo{title}{Order as an effect of disorder}.
\newblock \emph{\bibinfo{journal}{J. Phys. (France)}}
  \textbf{\bibinfo{volume}{41}}, \bibinfo{pages}{1263--1272}
  (\bibinfo{year}{1980}).

\bibitem{Shender1982}
\bibinfo{author}{Shender, E.}
\newblock \bibinfo{title}{Antiferromagnetic garnets with fluctuationally
  interacting sublattices}.
\newblock \emph{\bibinfo{journal}{Sov. Phys. JETP}}
  \textbf{\bibinfo{volume}{56}}, \bibinfo{pages}{178--184}
  (\bibinfo{year}{1982}).

\bibitem{Henley1989}
\bibinfo{author}{Henley, C.~L.}
\newblock \bibinfo{title}{Ordering due to disorder in a frustrated vector
  antiferromagnet}.
\newblock \emph{\bibinfo{journal}{Phys. Rev. Lett.}}
  \textbf{\bibinfo{volume}{62}}, \bibinfo{pages}{2056--2059}
  (\bibinfo{year}{1989}).

\bibitem{Lacroix2011}
\bibinfo{editor}{Lacroix, C.}, \bibinfo{editor}{Mendels, P.} \&
  \bibinfo{editor}{Mila, F.} (eds.) \emph{\bibinfo{title}{Introduction to
  Frustrated Magnetism: Materials, Experiments, Theory}}
  (\bibinfo{publisher}{Springer-Verlag}, \bibinfo{address}{Heidelberg},
  \bibinfo{year}{2011}).

\bibitem{Chubukov1991}
\bibinfo{author}{Chubukov, A.~V.} \& \bibinfo{author}{Golosov, D.~I.}
\newblock \bibinfo{title}{Quantum theory of an antiferromagnet on a triangular
  lattice in a magnetic field}.
\newblock \emph{\bibinfo{journal}{J. Phys.: Condens. Matter}}
  \textbf{\bibinfo{volume}{3}}, \bibinfo{pages}{69--82} (\bibinfo{year}{1991}).

\bibitem{Zhitomirsky2000}
\bibinfo{author}{Zhitomirsky, M.~E.}, \bibinfo{author}{Honecker, A.} \&
  \bibinfo{author}{Petrenko, O.~A.}
\newblock \bibinfo{title}{Field induced ordering in highly frustrated
  antiferromagnets}.
\newblock \emph{\bibinfo{journal}{Phys. Rev. Lett.}}
  \textbf{\bibinfo{volume}{85}}, \bibinfo{pages}{3269--3272}
  (\bibinfo{year}{2000}).

\bibitem{Hida2005}
\bibinfo{author}{Hida, K.} \& \bibinfo{author}{Affleck, I.}
\newblock \bibinfo{title}{Quantum vs classical magnetization plateaus of
  {S=1/2} frustrated heisenberg chains}.
\newblock \emph{\bibinfo{journal}{J. Phys. Soc. Jpn.}}
  \textbf{\bibinfo{volume}{74}}, \bibinfo{pages}{1849--1857}
  (\bibinfo{year}{2005}).

\bibitem{Lacroix2011C10}
\bibinfo{author}{Takigawa, M.} \& \bibinfo{author}{Mila, F.}
\newblock \emph{\bibinfo{title}{Magnetization Plateaus}},
  chap.~\bibinfo{chapter}{10} (\bibinfo{publisher}{Springer},
  \bibinfo{address}{Heidelberg}, \bibinfo{year}{2011}).

\bibitem{Seabra2016}
\bibinfo{author}{Seabra, L.}, \bibinfo{author}{Sindzingre, P.},
  \bibinfo{author}{Momoi, T.} \& \bibinfo{author}{Shannon, N.}
\newblock \bibinfo{title}{Novel phases in a square-lattice frustrated
  ferromagnet: $\frac{1}{3}$-magnetization plateau, helicoidal spin liquid, and
  vortex crystal}.
\newblock \emph{\bibinfo{journal}{Phys. Rev. B}} \textbf{\bibinfo{volume}{93}},
  \bibinfo{pages}{085132} (\bibinfo{year}{2016}).

\bibitem{Ye2017}
\bibinfo{author}{Ye, M.} \& \bibinfo{author}{Chubukov, A.~V.}
\newblock \bibinfo{title}{Half-magnetization plateau in a heisenberg
  antiferromagnet on a triangular lattice}.
\newblock \emph{\bibinfo{journal}{Phys. Rev. B}} \textbf{\bibinfo{volume}{96}},
  \bibinfo{pages}{140406} (\bibinfo{year}{2017}).

\bibitem{Rice2002}
\bibinfo{author}{Rice, T.~M.}
\newblock \bibinfo{title}{To condense or not to condense}.
\newblock \emph{\bibinfo{journal}{Science}} \textbf{\bibinfo{volume}{298}},
  \bibinfo{pages}{760--761} (\bibinfo{year}{2002}).

\bibitem{Giamarchi2008}
\bibinfo{author}{Giamarchi, T.}, \bibinfo{author}{Ruegg, C.} \&
  \bibinfo{author}{Tchernyshyov, O.}
\newblock \bibinfo{title}{Bose-einstein condensation in magnetic insulators}.
\newblock \emph{\bibinfo{journal}{Nat. Phys.}} \textbf{\bibinfo{volume}{4}},
  \bibinfo{pages}{198--204} (\bibinfo{year}{2008}).

\bibitem{Zapf2014}
\bibinfo{author}{Zapf, V.}, \bibinfo{author}{Jaime, M.} \&
  \bibinfo{author}{Batista, C.~D.}
\newblock \bibinfo{title}{Bose-einstein condensation in quantum magnets}.
\newblock \emph{\bibinfo{journal}{Rev. Mod. Phys.}}
  \textbf{\bibinfo{volume}{86}}, \bibinfo{pages}{563--614}
  (\bibinfo{year}{2014}).

\bibitem{Haravifard2016}
\bibinfo{author}{Haravifard, S.} \emph{et~al.}
\newblock \bibinfo{title}{Crystallization of spin superlattices with pressure
  and field in the layered magnet {SrCu$_2$(BO$_3$)$_2$}}.
\newblock \emph{\bibinfo{journal}{Nat. Comm.}} \textbf{\bibinfo{volume}{7}},
  \bibinfo{pages}{11956 EP} (\bibinfo{year}{2016}).

\bibitem{Honecker2004}
\bibinfo{author}{Honecker, A.}, \bibinfo{author}{Schulenburg, J.} \&
  \bibinfo{author}{Richter, J.}
\newblock \bibinfo{title}{Magnetization plateaus in frustrated
  antiferromagnetic quantum spin models}.
\newblock \emph{\bibinfo{journal}{J. Phys.: Condens. Matter}}
  \textbf{\bibinfo{volume}{16}}, \bibinfo{pages}{S749} (\bibinfo{year}{2004}).

\bibitem{Farnell2009}
\bibinfo{author}{Farnell, D. J.~J.}, \bibinfo{author}{Zinke, R.},
  \bibinfo{author}{Schulenburg, J.} \& \bibinfo{author}{Richter, J.}
\newblock \bibinfo{title}{High-order coupled cluster method study of frustrated
  and unfrustrated quantum magnets in external magnetic fields}.
\newblock \emph{\bibinfo{journal}{J. Phys.: Condens. Matter}}
  \textbf{\bibinfo{volume}{21}}, \bibinfo{pages}{406002}
  (\bibinfo{year}{2009}).

\bibitem{Sakai2011}
\bibinfo{author}{Sakai, T.} \& \bibinfo{author}{Nakano, H.}
\newblock \bibinfo{title}{Critical magnetization behavior of the triangular-
  and kagome-lattice quantum antiferromagnets}.
\newblock \emph{\bibinfo{journal}{Phys. Rev. B}} \textbf{\bibinfo{volume}{83}},
  \bibinfo{pages}{100405} (\bibinfo{year}{2011}).

\bibitem{Chen2013}
\bibinfo{author}{Chen, R.}, \bibinfo{author}{Ju, H.}, \bibinfo{author}{Jiang,
  H.-C.}, \bibinfo{author}{Starykh, O.~A.} \& \bibinfo{author}{Balents, L.}
\newblock \bibinfo{title}{Ground states of spin-$\frac{1}{2}$ triangular
  antiferromagnets in a magnetic field}.
\newblock \emph{\bibinfo{journal}{Phys. Rev. B}} \textbf{\bibinfo{volume}{87}},
  \bibinfo{pages}{165123} (\bibinfo{year}{2013}).

\bibitem{Yamamoto2014}
\bibinfo{author}{Yamamoto, D.}, \bibinfo{author}{Marmorini, G.} \&
  \bibinfo{author}{Danshita, I.}
\newblock \bibinfo{title}{Quantum phase diagram of the triangular-lattice
  {$XXZ$} model in a magnetic field}.
\newblock \emph{\bibinfo{journal}{Phys. Rev. Lett.}}
  \textbf{\bibinfo{volume}{112}}, \bibinfo{pages}{127203}
  (\bibinfo{year}{2014}).

\bibitem{Yamamoto2015}
\bibinfo{author}{Yamamoto, D.}, \bibinfo{author}{Marmorini, G.} \&
  \bibinfo{author}{Danshita, I.}
\newblock \bibinfo{title}{Microscopic model calculations for the magnetization
  process of layered triangular-lattice quantum antiferromagnets}.
\newblock \emph{\bibinfo{journal}{Phys. Rev. Lett.}}
  \textbf{\bibinfo{volume}{114}}, \bibinfo{pages}{027201}
  (\bibinfo{year}{2015}).

\bibitem{Nakano2017}
\bibinfo{author}{Nakano, H.} \& \bibinfo{author}{Sakai, T.}
\newblock \bibinfo{title}{Magnetization process of the spin-1/2
  triangular-lattice {Heisenberg} antiferromagnet with next-nearest-neighbor
  interactions---plateau or nonplateau---}.
\newblock \emph{\bibinfo{journal}{J. Phys. Soc. Jpn.}}
  \textbf{\bibinfo{volume}{86}}, \bibinfo{pages}{114705}
  (\bibinfo{year}{2017}).

\bibitem{Ono2003}
\bibinfo{author}{Ono, T.} \emph{et~al.}
\newblock \bibinfo{title}{Magnetization plateau in the frustrated quantum spin
  system {Cs$_2$CuBr$_4$}}.
\newblock \emph{\bibinfo{journal}{Phys. Rev. B}} \textbf{\bibinfo{volume}{67}},
  \bibinfo{pages}{104431} (\bibinfo{year}{2003}).

\bibitem{Ono2004}
\bibinfo{author}{Ono, T.} \emph{et~al.}
\newblock \bibinfo{title}{Magnetization plateaux of the {$S = 1/2$}
  two-dimensional frustrated antiferromagnet {Cs$_2$CuBr$_4$}}.
\newblock \emph{\bibinfo{journal}{J. Phys.: Condens. Matter}}
  \textbf{\bibinfo{volume}{16}}, \bibinfo{pages}{S773} (\bibinfo{year}{2004}).

\bibitem{Ono2005}
\bibinfo{author}{Ono, T.} \emph{et~al.}
\newblock \bibinfo{title}{Field-induced phase transitions driven by quantum
  fluctuation in {$S = 1/2$} anisotropic triangular antiferromagnet
  {Cs$_2$CuBr$_4$}}.
\newblock \emph{\bibinfo{journal}{Prog. Theor. Phys. Suppl.}}
  \textbf{\bibinfo{volume}{159}}, \bibinfo{pages}{217} (\bibinfo{year}{2005}).

\bibitem{Tsujii2007}
\bibinfo{author}{Tsujii, H.} \emph{et~al.}
\newblock \bibinfo{title}{Thermodynamics of the up-up-down phase of the
  {$S=\frac{1}{2}$} triangular-lattice antiferromagnet {Cs$_2$CuBr$_4$}}.
\newblock \emph{\bibinfo{journal}{Phys. Rev. B}} \textbf{\bibinfo{volume}{76}},
  \bibinfo{pages}{060406} (\bibinfo{year}{2007}).

\bibitem{Fortune2009}
\bibinfo{author}{Fortune, N.~A.} \emph{et~al.}
\newblock \bibinfo{title}{Cascade of magnetic-field-induced quantum phase
  transitions in a spin-$\frac{1}{2}$ triangular-lattice antiferromagnet}.
\newblock \emph{\bibinfo{journal}{Phys. Rev. Lett.}}
  \textbf{\bibinfo{volume}{102}}, \bibinfo{pages}{257201}
  (\bibinfo{year}{2009}).

\bibitem{Inami1996}
\bibinfo{author}{Inami, T.}, \bibinfo{author}{Ajiro, Y.} \&
  \bibinfo{author}{Goto, T.}
\newblock \bibinfo{title}{Magnetization process of the triangular lattice
  antiferromagnets, {RbFe(MoO$_4$)$_2$} and {CsFe(SO$_4$)$_2$}}.
\newblock \emph{\bibinfo{journal}{J. Phys. Soc. Jpn.}}
  \textbf{\bibinfo{volume}{65}}, \bibinfo{pages}{2374--2376}
  (\bibinfo{year}{1996}).

\bibitem{Svistov2003}
\bibinfo{author}{Svistov, L.~E.} \emph{et~al.}
\newblock \bibinfo{title}{Quasi-two-dimensional antiferromagnet on a triangular
  lattice {RbFe(MoO$_{4}$)$_2$}}.
\newblock \emph{\bibinfo{journal}{Phys. Rev. B}} \textbf{\bibinfo{volume}{67}},
  \bibinfo{pages}{094434} (\bibinfo{year}{2003}).

\bibitem{White2013}
\bibinfo{author}{White, J.~S.} \emph{et~al.}
\newblock \bibinfo{title}{Multiferroicity in the generic easy-plane triangular
  lattice antiferromagnet {RbFe(MoO${}_{4}$)${}_{2}$}}.
\newblock \emph{\bibinfo{journal}{Phys. Rev. B}} \textbf{\bibinfo{volume}{88}},
  \bibinfo{pages}{060409} (\bibinfo{year}{2013}).

\bibitem{Doi2004}
\bibinfo{author}{Doi, Y.}, \bibinfo{author}{Hinatsu, Y.} \&
  \bibinfo{author}{Ohoyama, K.}
\newblock \bibinfo{title}{Structural and magnetic properties of
  pseudo-two-dimensional triangular antiferromagnets {Ba$_3$$M$Sb$_2$O$_9$}
  ({$M$}^^e2^^80^^89= {Mn}, {Co}, and {Ni})}.
\newblock \emph{\bibinfo{journal}{J. Phys.: Condens. Matter}}
  \textbf{\bibinfo{volume}{16}}, \bibinfo{pages}{8923} (\bibinfo{year}{2004}).

\bibitem{Shirata2012}
\bibinfo{author}{Shirata, Y.}, \bibinfo{author}{Tanaka, H.},
  \bibinfo{author}{Matsuo, A.} \& \bibinfo{author}{Kindo, K.}
\newblock \bibinfo{title}{Experimental realization of a spin-$1/2$
  triangular-lattice {Heisenberg} antiferromagnet}.
\newblock \emph{\bibinfo{journal}{Phys. Rev. Lett.}}
  \textbf{\bibinfo{volume}{108}}, \bibinfo{pages}{057205}
  (\bibinfo{year}{2012}).

\bibitem{Zhou2012}
\bibinfo{author}{Zhou, H.~D.} \emph{et~al.}
\newblock \bibinfo{title}{Successive phase transitions and extended
  spin-excitation continuum in the {$S=\frac{1}{2}$} triangular-lattice
  antiferromagnet {Ba$_{3}$CoSb$_{2}$O$_{9}$}}.
\newblock \emph{\bibinfo{journal}{Phys. Rev. Lett.}}
  \textbf{\bibinfo{volume}{109}}, \bibinfo{pages}{267206}
  (\bibinfo{year}{2012}).

\bibitem{Susuki2013}
\bibinfo{author}{Susuki, T.} \emph{et~al.}
\newblock \bibinfo{title}{Magnetization process and collective excitations in
  the {$S=1/2$} triangular-lattice {Heisenberg} antiferromagnet
  {Ba$_{3}$CoSb$_{2}$O$_{9}$}}.
\newblock \emph{\bibinfo{journal}{Phys. Rev. Lett.}}
  \textbf{\bibinfo{volume}{110}}, \bibinfo{pages}{267201}
  (\bibinfo{year}{2013}).

\bibitem{Naruse2014}
\bibinfo{author}{Naruse, K.} \emph{et~al.}
\newblock \bibinfo{title}{Thermal conductivity in the triangular-lattice
  antiferromagnet {Ba$_3$CoSb$_2$O$_9$}}.
\newblock \emph{\bibinfo{journal}{J. Phys.: Conf. Series}}
  \textbf{\bibinfo{volume}{568}}, \bibinfo{pages}{042014}
  (\bibinfo{year}{2014}).

\bibitem{Koutroulakis2015}
\bibinfo{author}{Koutroulakis, G.} \emph{et~al.}
\newblock \bibinfo{title}{Quantum phase diagram of the ${S}=\frac{1}{2}$
  triangular-lattice antiferromagnet {Ba$_3$CoSb$_2$O$_9$}}.
\newblock \emph{\bibinfo{journal}{Phys. Rev. B}} \textbf{\bibinfo{volume}{91}},
  \bibinfo{pages}{024410} (\bibinfo{year}{2015}).

\bibitem{Quirion2015}
\bibinfo{author}{Quirion, G.} \emph{et~al.}
\newblock \bibinfo{title}{Magnetic phase diagram of {Ba$_3$CoSb$_2$O$_9$} as
  determined by ultrasound velocity measurements}.
\newblock \emph{\bibinfo{journal}{Phys. Rev. B}} \textbf{\bibinfo{volume}{92}},
  \bibinfo{pages}{014414} (\bibinfo{year}{2015}).

\bibitem{Ma2016}
\bibinfo{author}{Ma, J.} \emph{et~al.}
\newblock \bibinfo{title}{Static and dynamical properties of the spin-$1/2$
  equilateral triangular-lattice antiferromagnet {Ba$_3$CoSb$_2$O$_9$}}.
\newblock \emph{\bibinfo{journal}{Phys. Rev. Lett.}}
  \textbf{\bibinfo{volume}{116}}, \bibinfo{pages}{087201}
  (\bibinfo{year}{2016}).

\bibitem{Sera2016}
\bibinfo{author}{Sera, A.} \emph{et~al.}
\newblock \bibinfo{title}{${S}=\frac{1}{2}$ triangular-lattice antiferromagnets
  {Ba$_3$CoSb$_2$O$_9$} and {CsCuCl$_3$}: Role of spin-orbit coupling,
  crystalline electric field effect, and {Dzyaloshinskii-Moriya} interaction}.
\newblock \emph{\bibinfo{journal}{Phys. Rev. B}} \textbf{\bibinfo{volume}{94}},
  \bibinfo{pages}{214408} (\bibinfo{year}{2016}).

\bibitem{Ito2017}
\bibinfo{author}{Ito, S.} \emph{et~al.}
\newblock \bibinfo{title}{Structure of the magnetic excitations in the spin-1/2
  triangular-lattice heisenberg antiferromagnet {Ba$_3$CoSb$_2$O$_9$}}.
\newblock \emph{\bibinfo{journal}{Nat. Comm.}} \textbf{\bibinfo{volume}{8}},
  \bibinfo{pages}{235} (\bibinfo{year}{2017}).

\bibitem{Alicea2009}
\bibinfo{author}{Alicea, J.}, \bibinfo{author}{Chubukov, A.~V.} \&
  \bibinfo{author}{Starykh, O.~A.}
\newblock \bibinfo{title}{Quantum stabilization of the $1/3$-magnetization
  plateau in {Cs$_2$CuBr$_4$}}.
\newblock \emph{\bibinfo{journal}{Phys. Rev. Lett.}}
  \textbf{\bibinfo{volume}{102}}, \bibinfo{pages}{137201}
  (\bibinfo{year}{2009}).

\bibitem{Coletta2013}
\bibinfo{author}{Coletta, T.}, \bibinfo{author}{Zhitomirsky, M.~E.} \&
  \bibinfo{author}{Mila, F.}
\newblock \bibinfo{title}{Quantum stabilization of classically unstable plateau
  structures}.
\newblock \emph{\bibinfo{journal}{Phys. Rev. B}} \textbf{\bibinfo{volume}{87}},
  \bibinfo{pages}{060407} (\bibinfo{year}{2013}).

\bibitem{Coletta2016}
\bibinfo{author}{Coletta, T.}, \bibinfo{author}{T\'oth, T.~A.},
  \bibinfo{author}{Penc, K.} \& \bibinfo{author}{Mila, F.}
\newblock \bibinfo{title}{Semiclassical theory of the magnetization process of
  the triangular lattice heisenberg model}.
\newblock \emph{\bibinfo{journal}{Phys. Rev. B}} \textbf{\bibinfo{volume}{94}},
  \bibinfo{pages}{075136} (\bibinfo{year}{2016}).

\bibitem{Kawamura1985}
\bibinfo{author}{Kawamura, H.} \& \bibinfo{author}{Miyashita, S.}
\newblock \bibinfo{title}{Phase transition of the heisenberg antiferromagnet on
  the triangular lattice in a magnetic field}.
\newblock \emph{\bibinfo{journal}{J. Phys. Soc. Jpn.}}
  \textbf{\bibinfo{volume}{54}}, \bibinfo{pages}{4530--4538}
  (\bibinfo{year}{1985}).

\bibitem{Zhitomirsky2013}
\bibinfo{author}{Zhitomirsky, M.~E.} \& \bibinfo{author}{Chernyshev, A.~L.}
\newblock \bibinfo{title}{\textit{Colloquium} : Spontaneous magnon decays}.
\newblock \emph{\bibinfo{journal}{Rev. Mod. Phys.}}
  \textbf{\bibinfo{volume}{85}}, \bibinfo{pages}{219--242}
  (\bibinfo{year}{2013}).

\bibitem{Chernyshev2006}
\bibinfo{author}{Chernyshev, A.~L.} \& \bibinfo{author}{Zhitomirsky, M.~E.}
\newblock \bibinfo{title}{Magnon decay in noncollinear quantum
  antiferromagnets}.
\newblock \emph{\bibinfo{journal}{Phys. Rev. Lett.}}
  \textbf{\bibinfo{volume}{97}}, \bibinfo{pages}{207202}
  (\bibinfo{year}{2006}).

\bibitem{Starykh2006}
\bibinfo{author}{Starykh, O.~A.}, \bibinfo{author}{Chubukov, A.~V.} \&
  \bibinfo{author}{Abanov, A.~G.}
\newblock \bibinfo{title}{Flat spin-wave dispersion in a triangular
  antiferromagnet}.
\newblock \emph{\bibinfo{journal}{Phys. Rev. B}} \textbf{\bibinfo{volume}{74}},
  \bibinfo{pages}{180403} (\bibinfo{year}{2006}).

\bibitem{Chernyshev2009}
\bibinfo{author}{Chernyshev, A.~L.} \& \bibinfo{author}{Zhitomirsky, M.~E.}
\newblock \bibinfo{title}{Spin waves in a triangular lattice antiferromagnet:
  Decays, spectrum renormalization, and singularities}.
\newblock \emph{\bibinfo{journal}{Phys. Rev. B}} \textbf{\bibinfo{volume}{79}},
  \bibinfo{pages}{144416} (\bibinfo{year}{2009}).

\bibitem{Mourigal2013a}
\bibinfo{author}{Mourigal, M.}, \bibinfo{author}{Fuhrman, W.~T.},
  \bibinfo{author}{Chernyshev, A.~L.} \& \bibinfo{author}{Zhitomirsky, M.~E.}
\newblock \bibinfo{title}{Dynamical structure factor of the triangular-lattice
  antiferromagnet}.
\newblock \emph{\bibinfo{journal}{Phys. Rev. B}} \textbf{\bibinfo{volume}{88}},
  \bibinfo{pages}{094407} (\bibinfo{year}{2013}).

\bibitem{Paddison2016}
\bibinfo{author}{Paddison, J. A.~M.} \emph{et~al.}
\newblock \bibinfo{title}{Continuous excitations of the triangular-lattice
  quantum spin liquid {YbMgGaO$_4$}}.
\newblock \emph{\bibinfo{journal}{Nat. Phys.}} \textbf{\bibinfo{volume}{13}},
  \bibinfo{pages}{117 EP --} (\bibinfo{year}{2016}).

\bibitem{Park2016}
\bibinfo{author}{Park, K.} \emph{et~al.}
\newblock \bibinfo{title}{Magnon-phonon coupling and two-magnon continuum in
  the two-dimensional triangular antiferromagnet ${\mathrm{cucro}}_{2}$}.
\newblock \emph{\bibinfo{journal}{Phys. Rev. B}} \textbf{\bibinfo{volume}{94}},
  \bibinfo{pages}{104421} (\bibinfo{year}{2016}).

\bibitem{Christensen2007}
\bibinfo{author}{Christensen, N.~B.} \emph{et~al.}
\newblock \bibinfo{title}{Quantum dynamics and entanglement of spins on a
  square lattice}.
\newblock \emph{\bibinfo{journal}{Proc. Natl. Acad. Sci. USA}}
  \textbf{\bibinfo{volume}{104}}, \bibinfo{pages}{15264--15269}
  (\bibinfo{year}{2007}).

\bibitem{Tsyrulin2009}
\bibinfo{author}{Tsyrulin, N.} \emph{et~al.}
\newblock \bibinfo{title}{Quantum effects in a weakly frustrated ${S}=1/2$
  two-dimensional heisenberg antiferromagnet in an applied magnetic field}.
\newblock \emph{\bibinfo{journal}{Phys. Rev. Lett.}}
  \textbf{\bibinfo{volume}{102}}, \bibinfo{pages}{197201}
  (\bibinfo{year}{2009}).

\bibitem{Headings2010}
\bibinfo{author}{Headings, N.~S.}, \bibinfo{author}{Hayden, S.~M.},
  \bibinfo{author}{Coldea, R.} \& \bibinfo{author}{Perring, T.~G.}
\newblock \bibinfo{title}{Anomalous high-energy spin excitations in the
  high-${T}_{c}$ superconductor-parent antiferromagnet {La$_{2}$CuO$_{4}$}}.
\newblock \emph{\bibinfo{journal}{Phys. Rev. Lett.}}
  \textbf{\bibinfo{volume}{105}}, \bibinfo{pages}{247001}
  (\bibinfo{year}{2010}).

\bibitem{Plumb2014b}
\bibinfo{author}{Plumb, K.~W.}, \bibinfo{author}{Savici, A.~T.},
  \bibinfo{author}{Granroth, G.~E.}, \bibinfo{author}{Chou, F.~C.} \&
  \bibinfo{author}{Kim, Y.-J.}
\newblock \bibinfo{title}{High-energy continuum of magnetic excitations in the
  two-dimensional quantum antiferromagnet {Sr$_2$CuO$_2$Cl$_2$}}.
\newblock \emph{\bibinfo{journal}{Phys. Rev. B}} \textbf{\bibinfo{volume}{89}},
  \bibinfo{pages}{180410} (\bibinfo{year}{2014}).

\bibitem{Dalla-Piazza2015}
\bibinfo{author}{Dalla~Piazza, B.} \emph{et~al.}
\newblock \bibinfo{title}{Fractional excitations in the square-lattice quantum
  antiferromagnet}.
\newblock \emph{\bibinfo{journal}{Nat. Phys}} \textbf{\bibinfo{volume}{11}},
  \bibinfo{pages}{62--68} (\bibinfo{year}{2015}).

\bibitem{Powalski2015}
\bibinfo{author}{Powalski, M.}, \bibinfo{author}{Uhrig, G.~S.} \&
  \bibinfo{author}{Schmidt, K.~P.}
\newblock \bibinfo{title}{Roton minimum as a fingerprint of magnon-{Higgs}
  scattering in ordered quantum antiferromagnets}.
\newblock \emph{\bibinfo{journal}{Phys. Rev. Lett.}}
  \textbf{\bibinfo{volume}{115}}, \bibinfo{pages}{207202}
  (\bibinfo{year}{2015}).

\bibitem{Rodriguez-Carvajal1993}
\bibinfo{author}{Rodr{\'i}guez-Carvajal, J.}
\newblock \bibinfo{title}{Recent advances in magnetic structure determination
  by neutron powder diffraction}.
\newblock \emph{\bibinfo{journal}{Physica B Condens. Matter}}
  \textbf{\bibinfo{volume}{192}}, \bibinfo{pages}{55 -- 69}
  (\bibinfo{year}{1993}).

\bibitem{MACS}
\bibinfo{author}{Rodriguez, J.~A.} \emph{et~al.}
\newblock \bibinfo{title}{{MACS}---a new high intensity cold neutron
  spectrometer at {NIST}}.
\newblock \emph{\bibinfo{journal}{Meas. Sci. Technol.}}
  \textbf{\bibinfo{volume}{19}}, \bibinfo{pages}{034023}
  (\bibinfo{year}{2008}).

\bibitem{FLEXX}
\bibinfo{author}{Le, M.} \emph{et~al.}
\newblock \bibinfo{title}{Gains from the upgrade of the cold neutron
  triple-axis spectrometer {FLEXX} at the {BER-II} reactor}.
\newblock \emph{\bibinfo{journal}{Nucl. Instrum. Methods Phys. Res. Sect. A}}
  \textbf{\bibinfo{volume}{729}}, \bibinfo{pages}{220 -- 226}
  (\bibinfo{year}{2013}).

\end{thebibliography}

\section*{End Notes}
\subsection*{Acknowledgments}
We thank H.~Tanaka, A.~Chernyshev, and O.~Starykh for valuable discussions. J.M.~acknowledges the support of the Ministry of Science and Technology of China (2016YFA0300500). Y.K.~acknowledges the financial support by JSPS Grants-in-Aid for Scientific Research under Grant No.~JP16H02206. The work at Georgia Tech was supported by ORAU's Ralph E.~Powe Junior Faculty Enhancement Award (M.~Mourigal). H.D.Z.~acknowledges support from NSF-DMR-1350002. The work performed in NHMFL was supported by NSF-DMR-1157490 and the State of Florida. We are grateful for the access to the neutron beam time at the neutron facilities at NCNR, BER-II at Helmholtz-Zentrum Berlin and HFIR operated by ORNL. The research at HFIR at ORNL was sponsored by the Scientific User Facilities Division (T.H., H.B.C., and M.~Matsuda), Office of Basic Energy Sciences, U.S.~DOE. This work utilized MACS supported in part by the National Science Foundation under Agreement No.~DMR-1508249.

\subsection*{Author contributions}
Y.K.~and J.M.~conceived the project.
H.D.Z.~prepared the samples.
T.H., Y.Q., D.L.Q., Z.L., H.B.C., M.~Matsuda, L.G., M.~Mourigal, and J.M.~performed the neutron scattering experiments.
E.S.C.~measured the magnetization.
Y.K., L.G., C.D.B., and M.~Mourigal performed the NLSW calculations.
Y.K., J.M., M.~Mourigal, and C.D.B.~wrote the manuscript with comments from all the authors.

\subsection*{Competing financial interests}
The authors declare no competing financial interests.

\subsection*{Correspondence and materials}
Correspondence and requests for materials should be addressed to Y.K.~(yoshitomo.kamiya@riken.jp) and J.M.~(jma3@sjtu.edu.cn).

\end{document}